\documentclass[a4paper,12pt,twoside]{cpc-hepnp}

\usepackage{CJK,upgreek,fancyhdr}
\usepackage{graphicx}
\usepackage{amsmath}
\usepackage{color}

\begin{document}
\begin{CJK*}{GB}{gbsn}

\fancyhead[c]{\small Chinese Physics C~~~Vol. xx, No. x (201x) xxxxxx}
\fancyfoot[C]{\small 010201-\thepage}

\footnotetext[0]{Received 30 August 2016}

\title{On thermodynamic self-consistency of generic axiomatic-nonextensive statistics}

\author{Abdel Nasser Tawfik$^{1,2}$\email{a.tawfik@eng.mti.edu.eg},
\quad Hayam Yassin$^{3}$, 
\quad Eman R. Abo Elyazeed$^{3}$
}
\maketitle

\address{%
$^1$Egyptian Center for Theoretical Physics (ECTP), Modern University for Technology and Information (MTI), 11571 Cairo, Egypt\\
$^2$World Laboratory for Cosmology And Particle Physics (WLCAPP), 11571 Cairo, Egypt\\
$^3$Physics Department, Faculty of Women for Arts, Science and Education, Ain Shams University, 11577 Cairo, Egypt
}

\begin{abstract}
Generic axiomatic-nonextensive statistics characterized by two asymptotic properties, to each of them a scaling function is assigned, characterized by the exponents $c$ and $d$ for first and second scaling property, respectively, is formulated in a grand-canonical ensemble with finite volume in the thermodynamic limit. The thermodynamic properties of a relativistic ideal gas of hadron resonances are studied, analytically. It is found that this generic statistics satisfies the requirements of the equilibrium thermodynamics. Essential aspects of the thermodynamic self-consistency are clarified. Analytical expressions are proposed for the statistical fits of various transverse momentum distributions measured in most-central collisions at different collision energies and colliding systems. Estimations for the freezeout temperature ($T_{\mathtt{ch}}$) and the baryon chemical potential ($\mu_{\mathtt{b}}$) and the exponents $c$ and $d$ are determined. The earlier are found compatible with the parameters deduced from Boltzmann-Gibbs (BG) statistics (extensive), while the latter refer to generic nonextensivities. The resulting equivalence class $(c,d)$ is associated to stretched exponentials, where Lambert function reaches its asymptotic stability. In some measurements, the resulting nonextensive entropy is linearly composed on extensive entropies. Apart from power-scaling, the particle ratios and yields are excellent quantities to highlighting whether the particle production takes place, (non)extensively. Various particle ratios and yields measured by the STAR experiment in central collisions at $200$,  $62.4$ and $7.7~$GeV are fitted with this novel approach. We found that both $c$ and $d<1$, i.e. referring to neither BG- nor Tsallis-type statistics, but to $(c,d)$-entropy, where Lambert functions exponentially raise. The freezeout temperature and baryon chemical potential are found comparable with the ones deduced from BG statistics (extensive). We conclude that the particle production at STAR energies is likely a nonextensive process but not necessarily BG or Tsallis type.
\end{abstract}



\footnotetext[0]{\hspace*{-3mm}\raisebox{0.3ex}{$\scriptstyle\copyright$}2013
Chinese Physical Society and the Institute of High Energy Physics
of the Chinese Academy of Sciences and the Institute
of Modern Physics of the Chinese Academy of Sciences and IOP Publishing Ltd}%


\section{Introduction}
\label{sec:intro}

In theory of extensive (such as Boltzmann-Gibbs) and nonextensive (such as Tsallis) statistics, the thermodynamic consistency gives a phenomenological description for various phenomena in high-energy experiments \cite{Tawfik:2014eba,Tawfik:2010aq}. The earlier was utilized by Hagedorn \cite{Hagedorn1965} in 1960's to prove that the fireballs or heavy resonances lead to a bootstrap approach, i.e. further fireballs, which - in turn - consist of smaller fireballs and so on. The implementation of the nonextensive Tsallis statistics was introduced in Refs. \cite{ReFF1,ReFF2,Alberico:1999nh}. Assuming that the distribution function gets variations, due to possible symmetrical change, Tawfik applied nonextensive concepts to high-energy particle production \cite{Tawfik:2010uh}.
Recently, variouspapers were quite successful in explaining various aspects of the high-energy particle production using thermodynamically consistent nonextensive statistics of Tsallis-type \cite{Tripathy:2016hlg,Khuntia:2016ikm,Bhattacharyya:2015hya,Deppman:2012qt,Alberico:2005vu,Bhattacharyya:2015nwa,Wong2003}. These are based on the conjecture that replacing the Boltzmann factor by the $q$-exponential function of the Tsallis statistics, with $q>1$, leads to a good agreement with the experimental measurements at high energies. Recently, Tawfik explained that this method seems to fail to assure a full incorporation of nonextensivity because fluctuations, correlations, interactions among the produced particles besides the possible modification in the phase space of such an interacting system are not properly taken into account \cite{Tawfik:NICA1}. Again, the Tsallis distribution was widely applied to describe the hadron production \cite{Tripathy:2016hlg,Khuntia:2016ikm,Bhattacharyya:2015hya,Deppman:2012qt,Alberico:2005vu,Bhattacharyya:2015nwa,Wong2003}. At high transverse momentum spectra ($p_T$), some authors did not obtained power-law, while at low $p_T$, they obtained an approximate exponential-distribution.

In light of such a wide discrepancy and especially to find a unified statistical description for high-energy collisions, we introduce generic axiomatic-nonextensive statistics, in which the phase space determines the degree of (non)extensivity. The latter is not necessarily limited to extensive and/or intensive thermodynamic quantities, such as, temperature and baryon chemical potential. 

Regarding the success of Tsallis-statistics in describing the transverse momentum spectra, especially at high $p_T$, we first recall that Bialas claimed that such good fit would be incomplete \cite{bialas2015}. First, it is believed to ignore the contradiction between the applicability of such a statistical thermal approach at high energy and the perturbative QCD. The crucial question, how comes that the statistical thermal approach which describes well the lattice thermodynamics (non-perturbative) \cite{Tawfik:2014eba} should be assumed to do such claimed excellent job at ultra-relativistic energies, should be answered. Second, such a statistical fit seems to disbelieve the role of statistical cluster decay, which can be scaled as power laws very similar to that from Tsallis statistics. This simply means that the decay of statistical clusters is conjectured to be capable to explain the excellent reproduction of the measured transverse momenta rather than the Tsallis-type nonextensivity \cite{Tawfik:NICA1}. 

Also, Bialas \cite{bialas2015} presented within the statistical cluster-decay model a numerical analysis for the hadronization processes. It was found that the resulting transverse-momentum distribution can be a Tsallis-like one. Only in a very special case, where the fluctuations of the Lorentz factor and the temperature are given by Beta and Gamma distributions, respectively, the well-known Tsallis distribution can be obtained. The origin of these fluctuations was introduced in Ref. \cite{Wilka:2015}. Bialas explained that the produced hadronic clusters-decay, purely thermally, i.e. following Boltzmann-–Gibbs (BG) statistics \cite{bialas2015}. Furthermore, it is also supposed that the produced hadronic clusters move with a fluctuating Lorentz factor in the transverse direction, i.e. a power law. Thus, the production and the decay of such clusters would be be regarded as examples on superstatistics. The latter can be  understood as a kind of a superposition of two different types of statistics corresponding to nonequilibrium systems \cite{Beck:2003,Sattin:2006}. In other words, it was concluded \cite{Wilka:2015} that even superstatistics should be based on more general distributions than the Gamma type before being applied to multiparticle productions in high energy collisions. For the sake of completeness, we recall that the various power-law distributions have been already implemented to $pp$-collisions \cite{reFFF1,reFFF2,reFFF3,reFFF4,reFFF5}. 

The crucial question remains unanswered, namely, what is the origin and the degree of nonextensivity? And how to determine the degree of nonextensivity in a strongly correlated system as the relativists heavy-ion collisions? In the present work, we analyse the thermodynamic self-consistency of the generic axiomatic-nonextensive approach, which was introduced in \cite{Thurner1,Thurner2} and formulated for further implications in high-energy physics in \cite{Tawfik:NICA1,Tawfik:2015}.

In thermal equilibrium, the statistical mechanics is a thermodynamically self-consistent theory. It fulfils the requirements of the equilibrium thermodynamics, because  thermodynamic potential - in its thermodynamic limit - can be expressed as a first-order homogeneous function of extensive variables \cite{Parvan2015}. The entropy belongs to the fundamental checks for the thermodynamic self-consistency.  For a nonextensive quantum gas, the entropy should be fully constructed from bosons and fermions contributions; $S_q=S_q^{FD}+S_q^{BE}$, where $S_q^{FD}$ and $S_q^{BE}$ are Fermi-Dirac and Bose-Einstein nonextensive entropy, respectively \cite{Deppman:2012}.  In microcanonical \cite{Parvan20}, canonical \cite{Parvan21}, and grand-canonical \cite{J.Cleymans:2011} ensembles ensembles, the thermodynamic self-consistence of Tsallis statistics has been proved. If its entropic variable has extensive property, it was demonstrated that the homogeneity of the thermodynamic potential leads to the zeroth law of thermodynamics, i.e. additivity principle and Euler theorem. Also, the first and second laws of thermodynamics should be fulfilled. An additional ingredient of special importance in the particle production, namely the fireball self-consistency principle, should be guaranteed, as well.

The present paper is organized as follows. The generic axiomatic-nonextensive statistics is reviewed in section {\ref{sec:Generic}}.  The thermodynamic self-consistency will be discussed in section {\ref{sec:consistency}}. This is divided into nonextensive Boltzmann-Gibbs statistics (section {\ref{sec:Boltz}}) and generic axiomatic-nonextensive quantum statistics (section {\ref{sec:FD}}). Sections {\ref{sec:fitting}} and {\ref{sec:fittingPR}} are devoted to the fitting of the transverse momentum distributions and particle ratios and different beam energies and in different system sizes, respectively.  Section {\ref{sec:cncl}} elaborates the final conclusions.

\section{Short reminder to generic axiomatic-nonextensive statistics }
\label{sec:Generic}

Based on Hanel-Thurner entropy \cite{Thurner1,Thurner2,Tawfik:2015} which is fully expressed in Ref. \cite{Tawfik:NICA1},
\begin{eqnarray}
S_{c,d}[p] &=& \sum_{i=1}^{\Omega} {\cal A} \Gamma(d+1, 1 - c \log p_i) - {\cal B} \, p_i, \label{eq:NewExtns1}
\end{eqnarray}
where $p_i$ is the probability of $i$-th state and $\Gamma (a, b)=\int_{b}^{\infty}\, dt\, t^{a-1} \exp (-t)$ is incomplete gamma-function with ${\cal A}$ and ${\cal B}$ being arbitrary parameters, the generic axiomatic-nonextensive partition function for statistical processes in high-energy physics was suggested \cite{Tawfik:2015}. For a classical gas,
\begin{eqnarray}
\ln\, Z_{\mathtt{cl}}(T,\mu) &=& V\, \sum_i^{N_{\mathtt{M|B}}}\, g_i \int_0^{\infty} \frac{d^3\, {\bf p}}{(2\, \pi)^3}\; \varepsilon_{c,d,r}(x_i), \label{eq1}
\end{eqnarray}
where $V$ is the fireball volume and $x_i=\beta [\mu_{i}-E_i({\bf p})]$ with $E_i(p)=({\bf p}^2+m_i^2)^{1/2}$ being the dispersion relation of $i$-th state (particle). Straightforwardly, the quantum gas partition function reads
\begin{eqnarray}
\ln\, Z_{\mathtt{FB}}(T) &=& \pm V\, \sum_i^{N_{\mathtt{M|B}}}\, g_i \int_0^{\infty} \frac{d^3\, {\bf p}}{(2\, \pi)^3}\; \ln\left[1\pm\varepsilon_{c,d,r}(x_i)\right], \label{eq2}
\end{eqnarray}
where $\pm$ represent fermions (subscript $\mathtt{F}$) and bosons  (subscript $\mathtt{B}$), respectively. The distribution function $\varepsilon_{c,d,r}(x_i)$ is given as \cite{Thurner1,Thurner2}
\begin{equation}
\varepsilon_{c,d,r}(x)=\exp\left[ \frac{-d}{1-c} \left(W_k\left[B\left(1-\frac{x}{r}\right)^{\frac{1}{d}}\right]-W_k[B]\right)\right], \label{eq:epsln}
\end{equation}
where $W_k$ is Lambert W-function which has real solutions at $k=0$ with $d\geq 0$ and at $k=1$ with $d<0$,
\begin{equation}
B=\frac{(1-c)r}{1-(1-c)r} \exp\left[\frac{(1-c)r}{1-(1-c)r}\right],
\end{equation}
with $r=[1-c+c\,d]^{-1}$ and $c, d$ are two constants to be elaborated, shortly. Eq. \eqref{eq:epsln} is valid for both classical and quantum gas.

The equivalent classes $(c, d)$ stand for two exponents giving estimations for two scaling functions with two asymptotic properties \cite{Thurner1,Thurner2}. Statistical systems in their large size limit violating the fourth Shannon-Khinchin axiom are characterized by a unique pair of scaling exponents ($c,d$). Such systems have two asymptotic properties of their associated generalized entropies. Both properties are associated with one scaling function each. Each scaling function is characterized by one exponent; $c$ for first and $d$ for second property. These exponents define equivalence relations of entropic forms, i.e. two entropic forms are equivalent if their exponents are the same.

The various thermodynamic observables such as pressure ($p$), number ($n=N/V$), energy ($\epsilon=E/V$) and entropy density ($s=S/V$), respectively \cite{RAFELSKI} can be derived from Eq. \eqref{eq1} or Eq. \eqref{eq2}
\begin{eqnarray}
p = T \frac{\partial \ln Z}{\partial V}, \qquad
n = \frac{\partial p}{\partial \mu}, \qquad
\epsilon   = \frac{T^2}{V} \frac{\partial \ln Z}{\partial T} + \frac{\mu T}{V} \frac{\partial \ln Z}{\partial \mu}, \qquad
s = \frac{\partial p}{\partial T}. \label{4}
\end{eqnarray}

\section{Thermodynamic self-consistency}
\label{sec:consistency}

Here, we examine the thermodynamic self-consistency of generic axiomatic-nonextensive statistics in Boltzmann-Gibbs and quantum gases. The procedure goes as follows. We start with first and second laws of thermodynamic which control the system of interest. Then, we determine the thermodynamic properties of that system and confirm that both laws of thermodynamic are verified.

The first law of thermodynamics describes the change in energy ($dE$) in terms of a change in volume ($dV$) and entropy ($dS$):
\begin{equation}
dE(V,S)=-P\, dV + T\, dS, \label{firstLaw}
\end{equation}
where $T$ and $P$ are temperature and pressure coefficients, respectively. In the variation of the free energy, $F(V,T)=E - T\;S$, the dependence on the entropy as given in Eq. (\ref{firstLaw}) is to be replaced by a temperature-dependence
\begin{equation}
dF(V,T)= dE - T\, dS - S\, dT \equiv -P\, dV - S\, dT. \label{freeEnergy}
\end{equation}

If the system of interest contained a conserved number ($N$), it turns to be necessary to introduce chemical potential ($\mu$). Eqs. (\ref{firstLaw}) and (\ref{freeEnergy}) should be extended
\begin{eqnarray}
dE(V,S,N) &=& -P\, dV + T\, dS + \mu\, dN, \label{eq1} \\
dF(V,T,\mu) &=& -P\, dV - S\, dT - N\, d\mu. \label{eq2}
\end{eqnarray}
The pressure is derived as
\begin{equation}
p = -\left.\frac{\partial F}{\partial V}\right|_{T, \mu}.
\label{eq3}
\end{equation}
From Eqs. (\ref{eq1}), (\ref{eq2}) and (\ref{eq3}), thermodynamics relations can be deduced \cite{RAFELSKI}. Their justification proves the thermodynamic self-consistency,
\begin{eqnarray}
\textit{n} = \left.\frac{\partial \textit{p}}{\partial \mu}\right|_{T}, \qquad
T=\left. \frac{\partial \epsilon}{\partial \textit{s}}\right|_{\textit{n}}, \qquad
\textit{s} = \left. \frac{\partial \textit{p}}{\partial T}\right|_{\mu}, \qquad
\mu=\left.\frac{\partial \epsilon}{\partial \textit{n}}\right|_{\textit{s}}. \label{eq:thermoqnt1}
\end{eqnarray}
To verify the second law of thermodynamics one has to prove that $\partial s \geq 0$.

In the section that follows, all these thermodynamic quantities shall be derived from the generic axiomatic-nonextensive statistics for Boltzmann-Gibbs, Bose-Einstein and Fermi-Dirac statistical ensembles \cite{J.Cleymans:2011}.

\subsection{Nonextensive Boltzmann-Gibbs statistics}
\label{sec:Boltz}

As discussed in Ref. \cite{HT3}, the universality class $(c,d)$ is conjectured not only to characterize the entropy of the system of interest, entirely, but also to specify the distribution functions of that system in the thermodynamic limit. Thus, it is likely able to determine (non)extensivity of the system. For instance, if $(c,d)=(1,1)$, the system can be well described by BG statistics, while if $(c,d)=(q,0)$, the system possesses a Tsallis-type nonextensovity. Furthermore, if $(c,d)=(1,d)$,  stretched exponentials characterize that system. To our knowledge, further details about the physical meaning of ($c,d$) are being worked out by many colleagues. This would be published in the near future. The last case, for instance, requires that $d>0$ and $c\rightarrow 1$ so that $\lim_{c\rightarrow 1} \varepsilon_{c,d,r}(x) = \exp(-dr[1-x/r]^{1/d}-1)$. In light of this, for a single particle, the Boltzmann distribution in nonextensive system, Eq. \eqref{eq1} can be expressed as
\begin{eqnarray}
f(x)=\frac{1}{\varepsilon_{c,d,r}(x)},
\end{eqnarray}
from which various thermodynamic quantities can deduced
\begin{eqnarray}
\textit{p} &=& \frac{g T}{2 \pi^2} \int_0^\infty \textbf{p}^2 \ln\left[\varepsilon_{c,d,r}(x)\right] d\textbf{p}, \label{BGp}\\
\textit{n} &=& \frac{g}{2 \pi^2} \int_0^\infty \frac{\textbf{p}^2 \; W_0\left[B(1-\frac{x}{r})^{\frac{1}{d}}\right]}{(1-c) \left[r-x\right] \left(1+W_0\left[B(1-\frac{x}{r})^{\frac{1}{d}}\right]\right)} d\textbf{p}, \label{BGn}\\
\epsilon &=& \frac{-g}{2 \pi^2} \int_0^\infty \frac{\textbf{p}^2 (x\; T)\; W_0\left[B(1-\frac{x}{r})^{\frac{1}{d}}\right]}{(1-c) \left[r-x\right] \left(1+W_0\left[B(1-\frac{x}{r})^{\frac{1}{d}}\right]\right)} d\textbf{p} 
+ \frac{g\;\mu}{2 \pi^2} \int_0^\infty \frac{\textbf{p}^2 \; W_0\left[B(1-\frac{x}{r})^{\frac{1}{d}}\right]}{(1-c) \left[r-x\right] \left(1+W_0\left[B(1-\frac{x}{r})^{\frac{1}{d}}\right]\right)} d\textbf{p}, \label{BGE}\\
\textit{s} &=& \frac{g}{2 \pi^2} \int_0^\infty \textbf{p}^2 \ln\left[\varepsilon_{c,d,r}(x)\right] d\textbf{p} 
- \frac{g}{2\pi^2 T} \int_0^\infty \frac{\textbf{p}^2 \; (x\;T)\; W_0\left[B(1-\frac{x}{r})^{\frac{1}{d}}\right]}{(1-c)\left[r-x\right] \left(1+W_0\left[B(1-\frac{x}{r})^{\frac{1}{d}}\right]\right)} d\textbf{p}. \label{BGs1}
\end{eqnarray}
In an ideal gas approach, such as the hadron resonance gas (HRG) model, the thermodynamic quantities get contributions from each hadron resonance. Thus, a summation over different hadron resonances of which the statistical ensemble is consisting, should be added in front of the right-hand side.

To check and satisfy the thermodynamic self-consistency, let us rewrite Eq. (\ref{BGs1}) as
\begin{equation}
\textit{s} = \frac{\textit{p}}{T} +\frac{\epsilon}{T} - \frac{\mu\;\textit{n}}{T},
\label{s_B}
\end{equation}
and take its derivative with respect to $\epsilon$, then we get
\begin{equation}
\left. \frac{\partial \textit{s}}{\partial \epsilon}\right|_{\textit{n}}=\frac{1}{T}.
\label{first_proof}
\end{equation}
Furthermore, we derive from Eq. (\ref{eq1}) \cite{RAFELSKI},
\begin{equation}
\textit{p} = -\epsilon + T\, \textit{s} + \mu\, \textit{n}. \label{Consistency_start}
\end{equation}
At constant $T$, the derivative of pressure with respect to $\mu$ reads
\begin{equation}
\left.\frac{\partial \textit{p}}{\partial \mu}\right|_T =
-\frac{\partial \epsilon}{\partial \mu} +T\frac{\partial \textit{s}}{\partial \mu} + \textit{\textit{n}} + \mu\frac{\partial \textit{n}}{\partial \mu}. \label{Consistency1}
\end{equation}
The differentiations of Eqs. (\ref{BGn}), (\ref{BGE}), and (\ref{BGs1}) with respect to $\mu$ lead to
\begin{eqnarray}
\frac{\partial \textit{n}}{\partial \mu} &=& \frac{g}{2 \pi^2 T} \int_0^\infty \frac{\textbf{p}^2 \; W_0\left[B(1-\frac{x}{r})^{\frac{1}{d}}\right]^2}{(d-d\;c) r^2 \left(1-x/r\right)^2 \left(1+W_0\left[B(1-\frac{x}{r})^{\frac{1}{d}}\right]\right)^3} d\textbf{p} \nonumber \\
&-& \frac{g}{2 \pi^2 T} \int_0^\infty \frac{\textbf{p}^2 \; W_0\left[B(1-\frac{x}{r})^{\frac{1}{d}}\right]}{(d-d\;c) r^2 \left(1-x/r\right)^2 \left(1+W_0\left[B(1-\frac{x}{r})^{\frac{1}{d}}\right]\right)^2} d\textbf{p} \nonumber \\
&+&\frac{g}{2 \pi^2 T} \int_0^\infty \frac{\textbf{p}^2 \; W_0\left[B(1-\frac{x}{r})^{\frac{1}{d}}\right]}{(1-c) r^2 \left(1-x/r\right)^2 \left(1+W_0\left[B(1-\frac{x}{r})^{\frac{1}{d}}\right]\right)} d\textbf{p},\label{dn1}\\
\frac{\partial \epsilon}{\partial \mu} &=& \frac{g\;\mu}{2 \pi^2 T} \int_0^\infty \frac{\textbf{p}^2 \; \left(W_0\left[B(1-\frac{x}{r})^{\frac{1}{d}}\right]\right)^2}{(d-d\;c) r^2 \left(1-x/r\right)^2 \left(1+W_0\left[B(1-\frac{x}{r})^{\frac{1}{d}}\right]\right)^3} d\textbf{p} \nonumber \\
&-& \frac{g\;\mu}{2 \pi^2 T} \int_0^\infty \frac{\textbf{p}^2 \; W_0\left[B(1-\frac{x}{r})^{\frac{1}{d}}\right]}{(d-d\;c) r^2 \left(1-x/r\right)^2 \left(1+W_0\left[B(1-\frac{x}{r})^{\frac{1}{d}}\right]\right)^2} d\textbf{p} \nonumber \\
&+&\frac{g\;\mu}{2 \pi^2 T} \int_0^\infty \frac{\textbf{p}^2 \; W_0\left[B(1-\frac{x}{r})^{\frac{1}{d}}\right]}{(1-c) r^2 \left(1-x/r\right)^2 \left(1+W_0\left[B(1-\frac{x}{r})^{\frac{1}{d}}\right]\right)} d\textbf{p}\nonumber \\
&-& \frac{g}{2 \pi^2 T} \int_0^\infty \frac{\textbf{p}^2 \; (x\;T) \; W_0\left[B(1-\frac{x}{r})^{\frac{1}{d}}\right]^2}{(d-d\;c) r^2 \left(1-x/r\right)^2 \left(1+W_0\left[B(1-\frac{x}{r})^{\frac{1}{d}}\right]\right)^3} d\textbf{p} \nonumber \\
&+& \frac{g}{2 \pi^2 T} \int_0^\infty \frac{\textbf{p}^2 \; (x\;T) \; W_0\left[B(1-\frac{x}{r})^{\frac{1}{d}}\right]}{(d-d\;c)r^2 \left(1-x/r\right)^2 \left(1+W_0\left[B(1-\frac{x}{r})^{\frac{1}{d}}\right]\right)^2} d\textbf{p} \nonumber \\
&-&\frac{g}{2 \pi^2 T} \int_0^\infty \frac{\textbf{p}^2 \; (x\;T) \; W_0\left[B(1-\frac{x}{r})^{\frac{1}{d}}\right]}{(1-c) r^2 \left(1-x/r\right)^2 \left(1+W_0\left[B(1-\frac{x}{r})^{\frac{1}{d}}\right]\right)} d\textbf{p},\label{dE1}
\end{eqnarray}
\begin{eqnarray}
\frac{\partial \textit{s}}{\partial \mu} &=& \frac{-g}{2 \pi^2 T^2} \int_0^\infty \frac{\textbf{p}^2 \; (x\;T) \; W_0\left[B(1-\frac{x}{r})^{\frac{1}{d}}\right]^2}{(d-d\;c) r^2 \left(1-x/r\right)^2 \left(1+W_0\left[B(1-\frac{x}{r})^{\frac{1}{d}}\right]\right)^3} d\textbf{p} \nonumber \\
&+& \frac{g}{2 \pi^2 T^2} \int_0^\infty \frac{\textbf{p}^2\; (x\;T) \; W_0\left[B(1-\frac{x}{r})^{\frac{1}{d}}\right]}{(d-d\;c) r^2 \left(1-x/r\right)^2 \left(1+W_0\left[B(1-\frac{x}{r})^{\frac{1}{d}}\right]\right)^2} d\textbf{p} \nonumber \\
&-&\frac{g}{2 \pi^2 T^2} \int_0^\infty \frac{\textbf{p}^2 \; (x\;T) \; W_0\left[B(1-\frac{x}{r})^{\frac{1}{d}}\right]}{(1-c) r^2 \left(1-x/r\right)^2 \left(1+W_0\left[B(1-\frac{x}{r})^{\frac{1}{d}}\right]\right)} d\textbf{p}.\label{ds1}
\end{eqnarray}
Then, the substitution from Eqs. (\ref{BGn}), (\ref{dn1}), (\ref{dE1}) and (\ref{ds1}) into Eq. (\ref{Consistency1}) gives
\begin{equation}
\left.\frac{\partial \textit{p}}{\partial \mu}\right|_{T}=\textit{n}.
\label{second_proof}
\end{equation}

Also, from Eq. (\ref{s_B}) we can express the energy density as
\begin{equation}
\epsilon = T\;\textit{s} - \textit{p} + \mu\;\textit{n},
\end{equation}
Its differentiation with respect to $n$ is
\begin{equation}
\left. \frac{\partial \epsilon}{\partial \textit{n}}\right|_{\textit{s}} = - \frac{\partial \textit{p}}{\partial \textit{n}} + \mu + \textit{n}  \frac{\partial \mu}{\partial \textit{n}} =- \frac{\partial \textit{p}}{\partial \mu} \frac{\partial \mu}{\partial \textit{n}}+ \mu + \textit{n}  \frac{\partial \mu}{\partial n}.
\end{equation}
By substituting from Eq. (\ref{second_proof}) into this previous equation, we get
\begin{equation}
\left. \frac{\partial \epsilon}{\partial \textit{n}}\right|_{\textit{s}}=\mu,
\label{third_proof}
\end{equation}
which is a proof of the thermodynamic self-consistency.

To prove the fourth equation in Eq. (\ref{eq:thermoqnt1}), let us differentiate Eq. (\ref{Consistency_start}) with respect to temperature at constant $\mu$
\begin{equation}
\left.\frac{\partial \textit{p}}{\partial T}\right|_\mu = \textit{s} + T\frac{\partial \textit{s}}{\partial T}
-\frac{\partial \epsilon}{\partial T} + \mu \frac{\partial \textit{n}}{\partial T}, \label{Consistency2}
\end{equation}
The differentiations of Eqs. (\ref{BGn}), (\ref{BGE}) and (\ref{BGs1}) with respect to $T$ read
\begin{eqnarray}
\frac{\partial \textit{n}}{\partial T} &=& \frac{- g}{2 \pi^2 T^2} \int_0^\infty \frac{\textbf{p}^2 \; (x\;T)\; W_0\left[B(1-\frac{x}{r})^{\frac{1}{d}}\right]^2}{(d-d\;c) r^2 \left(1-x/r\right)^2 \left(1+W_0\left[B(1-\frac{x}{r})^{\frac{1}{d}}\right]\right)^3} d\textbf{p} \nonumber \\
&+& \frac{g}{2 \pi^2 T^2} \int_0^\infty \frac{\textbf{p}^2 \; (x\;T)\; W_0\left[B(1-\frac{x}{r})^{\frac{1}{d}}\right]}{(d-d\;c) r^2 \left(1-x/r\right)^2 \left(1+W_0\left[B(1-\frac{x}{r})^{\frac{1}{d}}\right]\right)^2} d\textbf{p} \nonumber \\
&-&\frac{g}{2 \pi^2 T^2} \int_0^\infty \frac{\textbf{p}^2 \; (x\;T)\; W_0\left[B(1-\frac{x}{r})^{\frac{1}{d}}\right]}{(1-c) r^2 \left(1-x/r\right)^2 \left(1+W_0\left[B(1-\frac{x}{r})^{\frac{1}{d}}\right]\right)} d\textbf{p},\label{dnT}
\end{eqnarray}
\begin{eqnarray}
\frac{\partial \epsilon}{\partial T} &=& \frac{g}{2 \pi^2 T^2} \int_0^\infty \frac{\textbf{p}^2 \; (x\;T)^2 \; W_0\left[B(1-\frac{x}{r})^{\frac{1}{d}}\right]^2}{(d-d\;c) r^2 \left(1-x/r\right)^2 \left(1+W_0\left[B(1-\frac{x}{r})^{\frac{1}{d}}\right]\right)^3} d\textbf{p} \nonumber \\
&-& \frac{g}{2 \pi^2 T^2} \int_0^\infty \frac{\textbf{p}^2 \; (x\;T)^2 \; W_0\left[B(1-\frac{x}{r})^{\frac{1}{d}}\right]}{(d-d\;c)r^2 \left(1-x/r\right)^2 \left(1+W_0\left[B(1-\frac{x}{r})^{\frac{1}{d}}\right]\right)^2} d\textbf{p} \nonumber \\
&+&\frac{g}{2 \pi^2 T^2} \int_0^\infty \frac{\textbf{p}^2 \; (x\;T)^2 \; W_0\left[B(1-\frac{x}{r})^{\frac{1}{d}}\right]}{(1-c) r^2 \left(1-x/r\right)^2 \left(1+W_0\left[B(1-\frac{x}{r})^{\frac{1}{d}}\right]\right)} d\textbf{p}\nonumber \\
&-&\frac{g\;\mu}{2 \pi^2 T^2} \int_0^\infty \frac{\textbf{p}^2 \; (x\;T)\; \left(W_0\left[B(1-\frac{x}{r})^{\frac{1}{d}}\right]\right)^2}{(d-d\;c) r^2 \left(1-x/r\right)^2 \left(1+W_0\left[B(1-\frac{x}{r})^{\frac{1}{d}}\right]\right)^3} d\textbf{p} \nonumber \\
&+& \frac{g\;\mu}{2 \pi^2 T^2} \int_0^\infty \frac{\textbf{p}^2 \; (x\;T)\; W_0\left[B(1-\frac{x}{r})^{\frac{1}{d}}\right]}{(d-d\;c) r^2 \left(1-x/r\right)^2 \left(1+W_0\left[B(1-\frac{x}{r})^{\frac{1}{d}}\right]\right)^2} d\textbf{p} \nonumber \\
&-&\frac{g\;\mu}{2 \pi^2 T^2} \int_0^\infty \frac{\textbf{p}^2 \; (x\;T)\; W_0\left[B(1-\frac{x}{r})^{\frac{1}{d}}\right]}{(1-c) r^2 \left(1-x/r\right)^2 \left(1+W_0\left[B(1-\frac{x}{r})^{\frac{1}{d}}\right]\right)} d\textbf{p},\label{dET} \\
\frac{\partial \textit{s}}{\partial T} &=& \frac{g}{2 \pi^2 T^3} \int_0^\infty \frac{\textbf{p}^2 \; (x\;T)^2 \; W_0\left[B(1-\frac{x}{r})^{\frac{1}{d}}\right]^2}{(d-d\;c) r^2 \left(1-x/r\right)^2 \left(1+W_0\left[B(1-\frac{x}{r})^{\frac{1}{d}}\right]\right)^3} d\textbf{p} \nonumber \\
&-& \frac{g}{2 \pi^2 T^3} \int_0^\infty \frac{\textbf{p}^2\; (x\;T)^2 \; W_0\left[B(1-\frac{x}{r})^{\frac{1}{d}}\right]}{(d-d\;c) r^2 \left(1-x/r\right)^2 \left(1+W_0\left[B(1-\frac{x}{r})^{\frac{1}{d}}\right]\right)^2} d\textbf{p} \nonumber \\
&+&\frac{g}{2 \pi^2 T^3} \int_0^\infty \frac{\textbf{p}^2 \; (x\;T)^2 \; W_0\left[B(1-\frac{x}{r})^{\frac{1}{d}}\right]}{(1-c) r^2 \left(1-x/r\right)^2 \left(1+W_0\left[B(1-\frac{x}{r})^{\frac{1}{d}}\right]\right)} d\textbf{p}.\label{dsT}
\end{eqnarray}
Then, the substitution from Eqs. (\ref{BGs1}), (\ref{dnT}), (\ref{dET}) and (\ref{dsT}) into Eq. (\ref{Consistency2}) leads to
\begin{equation}
\left.\frac{\partial \textit{p}}{\partial T}\right|_{\mu}=\textit{s}.
\label{fourth_proof}
\end{equation}
It is apparent that the given definitions of temperature, number density, chemical potential and entropy density are thermodynamically consistent.

The first law of thermodynamics describes the consequences of heat transfer, Eq. (\ref{s_B}), while the second law sets constrains on it, i.e. $\delta s \geq 0$,
\begin{eqnarray}
\partial s \simeq \frac{1}{T}\left[\partial \epsilon - \mu \partial n \right],
\end{eqnarray}
where $\partial \mu/\partial T$ is conjectured to vanish. Therefore, the second law of thermodynamics is fulfilled, if
\begin{eqnarray}
\partial \epsilon \geq \mu \partial n.
\end{eqnarray}
This inequality is fulfilled, when comparing Eq. (\ref{dnT}) with Eq. (\ref{dET}). Two cases can be classified. Firstly, $\partial \epsilon > \mu \partial n$ is obvious, at arbitrary $\mu$. Secondly, $\partial \epsilon = \mu \partial n$ is also obtained, at $c=1$ or $\mu=\epsilon$ or $\mu=\epsilon + T r$.

\subsection{Generic axiomatic-nonextensive quantum statistics}
\label{sec:FD}

For quantum statistics, the distribution function reads
\begin{eqnarray}
f(x)=\frac{1}{1\pm \varepsilon_{c,d,r}(x)},
\end{eqnarray}
Accordingly, the various thermodynamic quantities can be deduced
\begin{eqnarray}
\textit{p} &=& \frac{g T}{2 \pi^2} \int_0^\infty \textbf{p}^2 \ln\left[1\pm\varepsilon_{c,d,r}(x)\right] d\textbf{p}, \label{FDp}\\
\textit{n} &=& \frac{\pm g}{2 \pi^2} \int_0^\infty \frac{\textbf{p}^2 \; \varepsilon_{c,d,r}(x) \; W_0\left[B(1-\frac{x}{r})^{\frac{1}{d}}\right]}{(1-c)\left[1\pm\varepsilon_{c,d,r}(x)\right] \left(r-x\right) \left(1+W_0\left[B(1-\frac{x}{r})^{\frac{1}{d}}\right]\right)} d\textbf{p}, \label{FDn1}\\
\epsilon &=& \frac{\mp g}{2 \pi^2} \int_0 ^\infty \frac{\textbf{p}^2 \varepsilon_{c,d,r}(x)\; (x\;T) \; W_0\left[B(1-\frac{x}{r})^{\frac{1}{d}}\right]}{(1-c)\left[1\pm\varepsilon_{c,d,r}(x)\right] \left(r-x\right) \left(1+W_0\left[B(1-\frac{x}{r})^{\frac{1}{d}}\right]\right)} d\textbf{p} \nonumber \\
&\pm& \frac{g \mu}{2 \pi^2} \int_0 ^\infty \frac{\textbf{p}^2 \varepsilon_{c,d,r}(x)\; W_0\left[B(1-\frac{x}{r})^{\frac{1}{d}}\right]}{(1-c)\left[1\pm\varepsilon_{c,d,r}(x)\right] \left(r-x\right) \left(1+W_0\left[B(1-\frac{x}{r})^{\frac{1}{d}}\right]\right)} d\textbf{p}, \label{FDE}\\
\textit{s} &=& \frac{g}{2 \pi^2} \int_0^\infty \textbf{p}^2 \ln\left[1\pm \varepsilon_{c,d,r}(x)\right] d\textbf{p} 
\pm \frac{g}{2 \pi^2 T} \int_0^\infty \frac{\textbf{p}^2 \;\varepsilon_{c,d,r}(x) \; (x\;T) \; W_0\left[B(1-\frac{x}{r})^{\frac{1}{d}}\right]}{(1-c)\left[1\pm \varepsilon_{c,d,r}(x)\right] \left(r-x\right) \left(1+W_0\left[B(1-\frac{x}{r})^{\frac{1}{d}}\right]\right)} d\textbf{p}. \label{FDs1}
\end{eqnarray}

The differentiations of Eqs. (\ref{FDn1}), (\ref{FDE}) and (\ref{FDs1}) with respect to $\mu$ give
\begin{eqnarray}
\frac{\partial \textit{n}}{\partial \mu} &=& \frac{\pm g}{2 \pi^2 T} \int_0^\infty \frac{\textbf{p}^2 \;\varepsilon_{c,d,r}(x)\; W_0\left[B(1-\frac{x}{r})^{\frac{1}{d}}\right]^2}{(d-d\;c)\left[1\pm \varepsilon_{c,d,r}(x)\right] r^2 \left(1-x/r\right)^2 \left(1+W_0\left[B(1-\frac{x}{r})^{\frac{1}{d}}\right]\right)^3} d\textbf{p} \nonumber \\
&\mp& \frac{g}{2 \pi^2 T} \int_0^\infty \frac{\textbf{p}^2 \;\varepsilon_{c,d,r}(x) \; W_0\left[B(1-\frac{x}{r})^{\frac{1}{d}}\right]}{(d-d\;c)\left[1\pm \varepsilon_{c,d,r}(x)\right] r^2 \left(1-x/r\right)^2 \left(1+W_0\left[B(1-\frac{x}{r})^{\frac{1}{d}}\right]\right)^2} d\textbf{p} \nonumber \\
&-& \frac{g}{2 \pi^2 T} \int_0^\infty \frac{\textbf{p}^2 \;\zeta_{c,d,r}(x) \; W_0\left[B(1-\frac{x}{r})^{\frac{1}{d}}\right]^2}{(1-c)^2\left[1\pm \varepsilon_{c,d,r}(x)\right]^2 r^2 \left(1-x/r\right)^2 \left(1+W_0\left[B(1-\frac{x}{r})^{\frac{1}{d}}\right]\right)^2} d\textbf{p} \nonumber \\
&\pm& \frac{g}{2 \pi^2 T} \int_0^\infty \frac{\textbf{p}^2 \;\varepsilon_{c,d,r}(x) \; W_0\left[B(1-\frac{x}{r})^{\frac{1}{d}}\right]^2}{(1-c)^2\left[1\pm \varepsilon_{c,d,r}(x)\right] r^2 \left(1-x/r\right)^2 \left(1+W_0\left[B(1-\frac{x}{r})^{\frac{1}{d}}\right]\right)^2} d\textbf{p} \nonumber \\
&\pm&\frac{g}{2 \pi^2 T} \int_0^\infty \frac{\textbf{p}^2 \;\varepsilon_{c,d,r}(x)\; W_0\left[B(1-\frac{x}{r})^{\frac{1}{d}}\right]}{(1-c)\left[1\pm \varepsilon_{c,d,r}(x)\right] r^2 \left(1-x/r\right)^2 \left(1+W_0\left[B(1-\frac{x}{r})^{\frac{1}{d}}\right]\right)} d\textbf{p},\label{dn2}
\end{eqnarray}
where
\begin{equation}
\zeta_{c,d,r}(x)=\exp\left[ \frac{-2d}{1-c} \left(W_k\left[B\left(1-\frac{x}{r}\right)^{\frac{1}{d}}\right]-W_k[B]\right)\right].
\end{equation}

\begin{eqnarray}
\frac{\partial \epsilon}{\partial \mu} &=& \frac{\pm g\;\mu}{2 \pi^2 T} \int_0^\infty \frac{\textbf{p}^2 \;\varepsilon_{c,d,r}(x)\; \left(W_0\left[B(1-\frac{x}{r})^{\frac{1}{d}}\right]\right)^2}{(d-d\;c)\left[1\pm \varepsilon_{c,d,r}(x)\right] r^2 \left(1-x/r\right)^2 \left(1+W_0\left[B(1-\frac{x}{r})^{\frac{1}{d}}\right]\right)^3} d\textbf{p} \nonumber \\
&\mp& \frac{g\;\mu}{2 \pi^2 T} \int_0^\infty \frac{\textbf{p}^2 \;\varepsilon_{c,d,r}(x) \; W_0\left[B(1-\frac{x}{r})^{\frac{1}{d}}\right]}{(d-d\;c)\left[1\pm \varepsilon_{c,d,r}(x)\right] r^2 \left(1-x/r\right)^2 \left(1+W_0\left[B(1-\frac{x}{r})^{\frac{1}{d}}\right]\right)^2} d\textbf{p} \nonumber \\
&-& \frac{g\;\mu}{2 \pi^2 T} \int_0^\infty \frac{\textbf{p}^2 \;\zeta_{c,d,r}(x) \; W_0\left[B(1-\frac{x}{r})^{\frac{1}{d}}\right]^2}{(1-c)^2\left[1\pm \varepsilon_{c,d,r}(x)\right] r^2 \left(1-x/r\right)^2 \left(1+W_0\left[B(1-\frac{x}{r})^{\frac{1}{d}}\right]\right)^2} d\textbf{p} \nonumber \\
&\pm& \frac{g\;\mu}{2 \pi^2 T} \int_0^\infty \frac{\textbf{p}^2 \;\varepsilon_{c,d,r}(x) \; W_0\left[B(1-\frac{x}{r})^{\frac{1}{d}}\right]^2}{(1-c)^2\left[1\pm \varepsilon_{c,d,r}(x)\right] r^2 \left(1-x/r\right)^2 \left(1+W_0\left[B(1-\frac{x}{r})^{\frac{1}{d}}\right]\right)^2} d\textbf{p} \nonumber \\
&\pm&\frac{g\;\mu}{2 \pi^2 T} \int_0^\infty \frac{\textbf{p}^2 \;\varepsilon_{c,d,r}(x)\; W_0\left[B(1-\frac{x}{r})^{\frac{1}{d}}\right]}{(1-c)\left[1\pm \varepsilon_{c,d,r}(x)\right] r^2 \left(1-x/r\right)^2 \left(1+W_0\left[B(1-\frac{x}{r})^{\frac{1}{d}}\right]\right)} d\textbf{p}\nonumber \\
&\mp& \frac{g}{2 \pi^2 T} \int_0^\infty \frac{\textbf{p}^2 \;\varepsilon_{c,d,r}(x)\; (x\;T) \; W_0\left[B(1-\frac{x}{r})^{\frac{1}{d}}\right]^2}{(d-d\;c)\left[1\pm \varepsilon_{c,d,r}(x)\right] r^2 \left(1-x/r\right)^2 \left(1+W_0\left[B(1-\frac{x}{r})^{\frac{1}{d}}\right]\right)^3} d\textbf{p} \nonumber \\
&\pm& \frac{g}{2 \pi^2 T} \int_0^\infty \frac{\textbf{p}^2 \;\varepsilon_{c,d,r}(x) \; (x\;T) \; W_0\left[B(1-\frac{x}{r})^{\frac{1}{d}}\right]}{(d-d\;c)\left[1\pm \varepsilon_{c,d,r}(x)\right] r^2 \left(1-x/r\right)^2 \left(1+W_0\left[B(1-\frac{x}{r})^{\frac{1}{d}}\right]\right)^2} d\textbf{p} \nonumber \\
&+& \frac{g}{2 \pi^2 T} \int_0^\infty \frac{\textbf{p}^2 \;\zeta_{c,d,r}(x) \; (x\;T) \; W_0\left[B(1-\frac{x}{r})^{\frac{1}{d}}\right]^2}{(1-c)^2\left[1\pm \varepsilon_{c,d,r}(x)\right] r^2 \left(1-x/r\right)^2 \left(1+W_0\left[B(1-\frac{x}{r})^{\frac{1}{d}}\right]\right)^2} d\textbf{p} \nonumber \\
&\mp& \frac{g}{2 \pi^2 T} \int_0^\infty \frac{\textbf{p}^2 \;\varepsilon_{c,d,r}(x) \; (x\;T) \; W_0\left[B(1-\frac{x}{r})^{\frac{1}{d}}\right]^2}{(1-c)^2\left[1\pm \varepsilon_{c,d,r}(x)\right] r^2 \left(1-x/r\right)^2 \left(1+W_0\left[B(1-\frac{x}{r})^{\frac{1}{d}}\right]\right)^2} d\textbf{p} \nonumber \\
&\mp&\frac{g}{2 \pi^2 T} \int_0^\infty \frac{\textbf{p}^2 \;\varepsilon_{c,d,r}(x)\; (x\;T) \; W_0\left[B(1-\frac{x}{r})^{\frac{1}{d}}\right]}{(1-c)\left[1\pm \varepsilon_{c,d,r}(x)\right] r^2 \left(1-x/r\right)^2 \left(1+W_0\left[B(1-\frac{x}{r})^{\frac{1}{d}}\right]\right)} d\textbf{p},\label{dE2}
\end{eqnarray}

\begin{eqnarray}
\frac{\partial \textit{s}}{\partial \mu} &=& \frac{\mp g}{2 \pi^2 T^2} \int_0^\infty \frac{\textbf{p}^2 \;\varepsilon_{c,d,r}(x)\; (x\;T) \; W_0\left[B(1-\frac{x}{r})^{\frac{1}{d}}\right]^2}{(d-d\;c)\left[1\pm \varepsilon_{c,d,r}(x)\right] r^2 \left(1-x/r\right)^2 \left(1+W_0\left[B(1-\frac{x}{r})^{\frac{1}{d}}\right]\right)^3} d\textbf{p} \nonumber \\
&\pm& \frac{g}{2 \pi^2 T^2} \int_0^\infty \frac{\textbf{p}^2 \;\varepsilon_{c,d,r}(x) \; (x\;T) \; W_0\left[B(1-\frac{x}{r})^{\frac{1}{d}}\right]}{(d-d\;c)\left[1\pm \varepsilon_{c,d,r}(x)\right] r^2 \left(1-x/r\right)^2 \left(1+W_0\left[B(1-\frac{x}{r})^{\frac{1}{d}}\right]\right)^2} d\textbf{p} \nonumber \\
&+& \frac{g}{2 \pi^2 T^2} \int_0^\infty \frac{\textbf{p}^2 \;\zeta_{c,d,r}(x) \; (x\;T) \; W_0\left[B(1-\frac{x}{r})^{\frac{1}{d}}\right]^2}{(1-c)^2\left[1\pm \varepsilon_{c,d,r}(x)\right]^2 r^2 \left(1-x/r\right)^2 \left(1+W_0\left[B(1-\frac{x}{r})^{\frac{1}{d}}\right]\right)^2} d\textbf{p} \nonumber \\
&\mp& \frac{g}{2 \pi^2 T^2} \int_0^\infty \frac{\textbf{p}^2 \;\varepsilon_{c,d,r}(x) \; (x\;T) \; W_0\left[B(1-\frac{x}{r})^{\frac{1}{d}}\right]^2}{(1-c)^2\left[1\pm \varepsilon_{c,d,r}(x)\right] r^2 \left(1-x/r\right)^2 \left(1+W_0\left[B(1-\frac{x}{r})^{\frac{1}{d}}\right]\right)^2} d\textbf{p} \nonumber \\
&\mp&\frac{g}{2 \pi^2 T^2} \int_0^\infty \frac{\textbf{p}^2 \;\varepsilon_{c,d,r}(x)\; (x\;T) \; W_0\left[B(1-\frac{x}{r})^{\frac{1}{d}}\right]}{(1-c)\left[1\pm \varepsilon_{c,d,r}(x)\right] r^2 \left(1-x/r\right)^2 \left(1+W_0\left[B(1-\frac{x}{r})^{\frac{1}{d}}\right]\right)} d\textbf{p},\label{ds2}
\end{eqnarray}

From the substitution from Eqs. (\ref{dn2}), (\ref{dE2}), (\ref{ds2}) and  (\ref{FDn}) into Eq. (\ref{Consistency1}), we get
\begin{equation}
\left.\frac{\partial \textit{p}}{\partial \mu}\right|_{T}=\textit{n}.
\label{first_proof2}
\end{equation}
Also, from Eqs. (\ref{FDE}) and (\ref{FDs1}), we get a direct relation between entropy and energy density,
\begin{equation}
\textit{s} = \frac{\textit{p}}{T} + \frac{\epsilon}{T} - \frac{\mu\;\textit{n}}{T}.
\label{s_Q}
\end{equation}
Thus, at constant $n$ and $p$  the differentiation of entropy density with respect to energy density leads to
\begin{equation}
\frac{\partial \textit{s}}{\partial \epsilon}=\frac{1}{T},
\label{second_proof2}
\end{equation}
which proves the thermodynamic consistency, Eqs. (\ref{eq:thermoqnt1}).

Also, from Eq. (\ref{s_Q}) we can write another form of energy density
\begin{equation}
\epsilon = T\;\textit{s} - \textit{p} + \mu\;\textit{n}.
\end{equation}
By differentiating the energy density with respect to $n$, we get
\begin{equation}
\left. \frac{\partial \epsilon}{\partial n}\right|_{\textit{s}}= - \frac{\partial \textit{p}}{\partial \textit{n}} + \mu + \textit{n}  \frac{\partial \mu}{\partial \textit{n}} = - \frac{\partial \textit{p}}{\partial \mu} \frac{\partial \mu}{\partial \textit{n}}+ \mu + \textit{n}  \frac{\partial \mu}{\partial \textit{n}}.
\end{equation}
Then, by substituting from Eq. (\ref{first_proof2}) into the previous equation,
\begin{equation}
\left. \frac{\partial \epsilon}{\partial \textit{n}}\right|_{\textit{s}}=\mu,
\label{third_proof2}
\end{equation}
which is a proof of the thermodynamic self-consistency.

To prove the fourth equation in Eq. (\ref{eq:thermoqnt1}), the differentiations of Eqs. (\ref{FDn1}), (\ref{FDE}) and (\ref{FDs1}) with respect to $T$ read
\begin{eqnarray}
\frac{\partial \textit{n}}{\partial T} &=& \frac{- g}{2 \pi^2 T^2} \int_0^\infty \frac{\textbf{p}^2 \;\varepsilon_{c,d,r}(x)\;(x\;T) \; W_0\left[B(1-\frac{x}{r})^{\frac{1}{d}}\right]^2}{(d-d\;c)\left[1\pm \varepsilon_{c,d,r}(x)\right] r^2 \left(1-x/r\right)^2 \left(1+W_0\left[B(1-\frac{x}{r})^{\frac{1}{d}}\right]\right)^3} d\textbf{p} \nonumber \\
&\pm& \frac{g}{2 \pi^2 T^2} \int_0^\infty \frac{\textbf{p}^2 \;\varepsilon_{c,d,r}(x) \; (x\;T) \; W_0\left[B(1-\frac{x}{r})^{\frac{1}{d}}\right]}{(d-d\;c)\left[1\pm \varepsilon_{c,d,r}(x)\right] r^2 \left(1-x/r\right)^2 \left(1+W_0\left[B(1-\frac{x}{r})^{\frac{1}{d}}\right]\right)^2} d\textbf{p} \nonumber \\
&+& \frac{g}{2 \pi^2 T^2} \int_0^\infty \frac{\textbf{p}^2 \;\zeta_{c,d,r}(x) \; (x\;T) \; W_0\left[B(1-\frac{x}{r})^{\frac{1}{d}}\right]^2}{(1-c)^2\left[1\pm \varepsilon_{c,d,r}(x)\right]^2 r^2 \left(1-x/r\right)^2 \left(1+W_0\left[B(1-\frac{x}{r})^{\frac{1}{d}}\right]\right)^2} d\textbf{p} \nonumber \\
&\mp& \frac{g}{2 \pi^2 T^2} \int_0^\infty \frac{\textbf{p}^2 \;\varepsilon_{c,d,r}(x) \; (x\;T) \; W_0\left[B(1-\frac{x}{r})^{\frac{1}{d}}\right]^2}{(1-c)^2\left[1\pm \varepsilon_{c,d,r}(x)\right] r^2 \left(1-x/r\right)^2 \left(1+W_0\left[B(1-\frac{x}{r})^{\frac{1}{d}}\right]\right)^2} d\textbf{p} \nonumber \\
&\mp&\frac{g}{2 \pi^2 T^2} \int_0^\infty \frac{\textbf{p}^2 \;\varepsilon_{c,d,r}(x)\; (x\;T) \; W_0\left[B(1-\frac{x}{r})^{\frac{1}{d}}\right]}{(1-c)\left[1\pm \varepsilon_{c,d,r}(x)\right] r^2 \left(1-x/r\right)^2 \left(1+W_0\left[B(1-\frac{x}{r})^{\frac{1}{d}}\right]\right)} d\textbf{p},\label{dnT2}
\end{eqnarray}

\begin{eqnarray}
\frac{\partial \epsilon}{\partial T} &=& \frac{\pm g}{2 \pi^2 T^2} \int_0^\infty \frac{\textbf{p}^2 \;\varepsilon_{c,d,r}(x)\; (x\;T)^2 \; W_0\left[B(1-\frac{x}{r})^{\frac{1}{d}}\right]^2}{(d-d\;c)\left[1\pm \varepsilon_{c,d,r}(x)\right] r^2 \left(1-x/r\right)^2 \left(1+W_0\left[B(1-\frac{x}{r})^{\frac{1}{d}}\right]\right)^3} d\textbf{p} \nonumber \\
&\mp& \frac{g}{2 \pi^2 T^2} \int_0^\infty \frac{\textbf{p}^2 \;\varepsilon_{c,d,r}(x) \; (x\;T)^2 \; W_0\left[B(1-\frac{x}{r})^{\frac{1}{d}}\right]}{(d-d\;c)\left[1\pm \varepsilon_{c,d,r}(x)\right] r^2 \left(1-x/r\right)^2 \left(1+W_0\left[B(1-\frac{x}{r})^{\frac{1}{d}}\right]\right)^2} d\textbf{p} \nonumber \\
&-& \frac{g}{2 \pi^2 T} \int_0^\infty \frac{\textbf{p}^2 \;\zeta_{c,d,r}(x) \; (x\;T)^2 \; W_0\left[B(1-\frac{x}{r})^{\frac{1}{d}}\right]^2}{(1-c)^2\left[1\pm \varepsilon_{c,d,r}(x)\right] r^2 \left(1-x/r\right)^2 \left(1+W_0\left[B(1-\frac{x}{r})^{\frac{1}{d}}\right]\right)^2} d\textbf{p} \nonumber \\
&\pm& \frac{g}{2 \pi^2 T^2} \int_0^\infty \frac{\textbf{p}^2 \;\varepsilon_{c,d,r}(x) \; (x\;T)^2 \; W_0\left[B(1-\frac{x}{r})^{\frac{1}{d}}\right]^2}{(1-c)^2\left[1\pm \varepsilon_{c,d,r}(x)\right] r^2 \left(1-x/r\right)^2 \left(1+W_0\left[B(1-\frac{x}{r})^{\frac{1}{d}}\right]\right)^2} d\textbf{p} \nonumber \\
&\pm&\frac{g}{2 \pi^2 T^2} \int_0^\infty \frac{\textbf{p}^2 \;\varepsilon_{c,d,r}(x)\; (x\;T)^2 \; W_0\left[B(1-\frac{x}{r})^{\frac{1}{d}}\right]}{(1-c)\left[1\pm \varepsilon_{c,d,r}(x)\right] r^2 \left(1-x/r\right)^2 \left(1+W_0\left[B(1-\frac{x}{r})^{\frac{1}{d}}\right]\right)} d\textbf{p}\nonumber \\
&\mp&\frac{g\;\mu}{2 \pi^2 T^2} \int_0^\infty \frac{\textbf{p}^2 \;\varepsilon_{c,d,r}(x)\; (x\;T)\; \left(W_0\left[B(1-\frac{x}{r})^{\frac{1}{d}}\right]\right)^2}{(d-d\;c)\left[1\pm \varepsilon_{c,d,r}(x)\right] r^2 \left(1-x/r\right)^2 \left(1+W_0\left[B(1-\frac{x}{r})^{\frac{1}{d}}\right]\right)^3} d\textbf{p} \nonumber \\
&\pm& \frac{g\;\mu}{2 \pi^2 T^2} \int_0^\infty \frac{\textbf{p}^2 \;\varepsilon_{c,d,r}(x) \; (x\;T)\; W_0\left[B(1-\frac{x}{r})^{\frac{1}{d}}\right]}{(d-d\;c)\left[1\pm \varepsilon_{c,d,r}(x)\right] r^2 \left(1-x/r\right)^2 \left(1+W_0\left[B(1-\frac{x}{r})^{\frac{1}{d}}\right]\right)^2} d\textbf{p} \nonumber \\
&+& \frac{g\;\mu}{2 \pi^2 T^2} \int_0^\infty \frac{\textbf{p}^2 \;\zeta_{c,d,r}(x) \; (x\;T)\; W_0\left[B(1-\frac{x}{r})^{\frac{1}{d}}\right]^2}{(1-c)^2\left[1\pm \varepsilon_{c,d,r}(x)\right] r^2 \left(1-x/r\right)^2 \left(1+W_0\left[B(1-\frac{x}{r})^{\frac{1}{d}}\right]\right)^2} d\textbf{p} \nonumber \\
&\mp& \frac{g\;\mu}{2 \pi^2 T^2} \int_0^\infty \frac{\textbf{p}^2 \;\varepsilon_{c,d,r}(x) \; (x\;T)\; W_0\left[B(1-\frac{x}{r})^{\frac{1}{d}}\right]^2}{(1-c)^2\left[1\pm \varepsilon_{c,d,r}(x)\right] r^2 \left(1-x/r\right)^2 \left(1+W_0\left[B(1-\frac{x}{r})^{\frac{1}{d}}\right]\right)^2} d\textbf{p} \nonumber \\
&\mp&\frac{g\;\mu}{2 \pi^2 T^2} \int_0^\infty \frac{\textbf{p}^2 \;\varepsilon_{c,d,r}(x)\; (x\;T)\; W_0\left[B(1-\frac{x}{r})^{\frac{1}{d}}\right]}{(1-c)\left[1\pm \varepsilon_{c,d,r}(x)\right] r^2 \left(1-x/r\right)^2 \left(1+W_0\left[B(1-\frac{x}{r})^{\frac{1}{d}}\right]\right)} d\textbf{p},\label{dET2}
\end{eqnarray}

\begin{eqnarray}
\frac{\partial \textit{s}}{\partial T} &=& \frac{\pm g}{2 \pi^2 T^3} \int_0^\infty \frac{\textbf{p}^2 \;\varepsilon_{c,d,r}(x)\; (x\;T)^2 \; W_0\left[B(1-\frac{x}{r})^{\frac{1}{d}}\right]^2}{(d-d\;c)\left[1\pm \varepsilon_{c,d,r}(x)\right] r^2 \left(1-x/r\right)^2 \left(1+W_0\left[B(1-\frac{x}{r})^{\frac{1}{d}}\right]\right)^3} d\textbf{p} \nonumber \\
&\mp& \frac{g}{2 \pi^2 T^3} \int_0^\infty \frac{\textbf{p}^2 \;\varepsilon_{c,d,r}(x) \; (x\;T)^2 \; W_0\left[B(1-\frac{x}{r})^{\frac{1}{d}}\right]}{(d-d\;c)\left[1\pm \varepsilon_{c,d,r}(x)\right] r^2 \left(1-x/r\right)^2 \left(1+W_0\left[B(1-\frac{x}{r})^{\frac{1}{d}}\right]\right)^2} d\textbf{p} \nonumber \\
&-& \frac{g}{2 \pi^2 T^3} \int_0^\infty \frac{\textbf{p}^2 \;\zeta_{c,d,r}(x) \; (x\;T)^2 \; W_0\left[B(1-\frac{x}{r})^{\frac{1}{d}}\right]^2}{(1-c)^2\left[1\pm \varepsilon_{c,d,r}(x)\right]^2 r^2 \left(1-x/r\right)^2 \left(1+W_0\left[B(1-\frac{x}{r})^{\frac{1}{d}}\right]\right)^2} d\textbf{p} \nonumber \\
&\pm& \frac{g}{2 \pi^2 T^3} \int_0^\infty \frac{\textbf{p}^2 \;\varepsilon_{c,d,r}(x) \; (x\;T)^2 \; W_0\left[B(1-\frac{x}{r})^{\frac{1}{d}}\right]^2}{(1-c)^2\left[1\pm \varepsilon_{c,d,r}(x)\right] r^2 \left(1-x/r\right)^2 \left(1+W_0\left[B(1-\frac{x}{r})^{\frac{1}{d}}\right]\right)^2} d\textbf{p} \nonumber \\
&\pm&\frac{g}{2 \pi^2 T^3} \int_0^\infty \frac{\textbf{p}^2 \;\varepsilon_{c,d,r}(x)\; (x\;T)^2 \; W_0\left[B(1-\frac{x}{r})^{\frac{1}{d}}\right]}{(1-c)\left[1\pm \varepsilon_{c,d,r}(x)\right] r^2 \left(1-x/r\right)^2 \left(1+W_0\left[B(1-\frac{x}{r})^{\frac{1}{d}}\right]\right)} d\textbf{p}.\label{dsT2}
\end{eqnarray}
Then, the substitution from Eqs. (\ref{FDs1}), (\ref{dnT2}), (\ref{dET2}) and (\ref{dsT2}) into Eq. (\ref{Consistency2}) gives
\begin{equation}
\left.\frac{\partial p}{\partial T}\right|_{\mu}=\textit{s}.
\label{fourth_proof2}
\end{equation}
So far, we have proved that the definitions of temperature, number density, chemical potential and entropy density within our formalism for nonextensive quantum statistics lead to expressions which satisfy first law of thermodynamics.

\section{Confronting our calculations to various experimental results}
\label{sec:fitting}

The statistical-thermal models have been successful in reproducing particle rations and yields at different energies \cite{Tawfik:2014eba,Cleymans:2005,Andronic:2006,Becattini:2005,Tripathy:2016hlg,Khuntia:2016ikm,Bhattacharyya:2015hya}. In these models, the hadronic phase can be modelled at chemical and thermal equilibrium. Accordingly, fitting parameters can then be identified. They construct a set of various thermal parameters. The most significant ones are the chemical freeze-out temperature and baryon chemical potential \cite{Tawfik:2014eba}.

The deconfined phase is dominated by quarks and gluons degrees-of-freedom. Most of their information can not be recognized due to the nature of such partonic QCD matter. While the integrated particle-yields are successfully constructed in the final state, the transverse momentum distribution should be described by a combination of transverse flow and statistical distributions (particle ratios and yields). In other words, the latter contains contributions from earlier stages of the collision, while the particle ratios and yields are conjectured to be fixed during the chemical and thermal equilibrium stages. Thus, we plan to confront this new generic approach to both types of experimental results, namely, transverse momentum spectra and particle ratios and yields. This shall introduced in the sections that follow. It intends to prove whether the proposed approach is indeed able to reflect the statistical nature of the system of interest.

\subsection{Transverse momentum distributions}

The particle distributions at large transverse momenta are indeed very interesting phenomena in high-energy particle production, but as discussed by Bialas \cite{bialas2015}, the applicability of the statistical-thermal models in this regime of the transverse momenta is debatable. These models, either extensive or nonextensive, are excellent approaches at low $p_T$. The scope of this paper is the examination of the thermodynamic self-consistency and then implementing the proposed approach in characterizing both transverse momentum distributions and particle ratios and yields. 

As discussed in the introduction, the hadronic clusters are assumed to undergo thermal decays, but simultaneously move in the transverse direction with a fluctuation Lorentz factor \cite{bialas2015}. Recently, this statistical cluster-decay model was implemented in high-energy physics \cite{Wilka:2015} and concluded that the well-known Tsallis distribution can be obtained in a very special case, namely, the fluctuations of the Lorentz factor and the relativistic temperature are given by Beta and Gamma distributions, respectively. The role of the statistical cluster-decay and its possible connections with the approach proposed in the present work shall be subjects of future works. 

We first implement the generic nonextensive statistical approach \cite{Tawfik:2015} in order to reproduce various transverse momentum distributions measured in different experiments \cite{UA1,reFFF4,Adare:2013esx,Abelev:2008ab,Abelev:2013haa,ALICE}. For quantum statistics, the total number of particles can be determined from Eq. (\ref{FDn})
\begin{eqnarray}
N &=& \pm\frac{V}{8 \pi^3} \sum_i g_i \int_0^\infty \frac{\textbf{p}^2 \; \varepsilon_{c,d,r}(x_i) \; W_0\left[B(1-\frac{x_i}{r})^{\frac{1}{d}}\right]}{(1-c)\left[1\pm\varepsilon_{c,d,r}(x_i)\right] \left(r-x_i\right) \left(1+W_0\left[B(1-\frac{x_i}{r})^{\frac{1}{d}}\right]\right)} d^3\textbf{p}, \label{FDn}
\end{eqnarray}
where $x_i=(\mu-E_i)/T$, $i \in [\mathtt{\pi}^+, \mathtt{K}^+, \mathtt{p}]$, and $r=[1-c+c\,d]^{-1}$.  Their degeneracy factors read $g_{\mathtt{\pi}^+}=g_{\mathtt{K}^+}=g_{\mathtt{p}}=1$. For antiparticles, $\mu$ is replaced by $-\mu$.

The corresponding momentum distribution for particles (or antiparticles) is given as
\begin{eqnarray}
\frac{1}{2 \pi}\frac{E d^3N}{d^3p}&=&\pm \frac{V}{8 \pi^3} \sum_i g_i E_i T \frac{\varepsilon_{c,d,r}\left(x_i\right) \; W_0\left[B\left(1-\frac{x_i}{r}\right)^{\frac{1}{d}}\right]}{(1-c)\left[1\pm\varepsilon_{c,d,r}\left(x_i\right)\right] \left(r T -\mu+E_i\right) \left(1+W_0\left[B\left(1-\frac{x_i}{r}\right)^{\frac{1}{d}}\right]\right)},
\end{eqnarray}
where $\varepsilon_{c,d,r}\left(x_i\right)$ is defined in Eq. (\ref{eq:epsln}).
In terms of rapidity ($y$) and transverse mass (${m_T}_i = \sqrt{p_T^2 + m^2_i}$), the transverse momentum distribution can be given as
\begin{eqnarray}
\frac{1}{2 \pi}\frac{d^2 N}{dy m_T dm_T}&=&\pm\frac{V}{8 \pi^3}\sum_i g_i T {m_T}_i \cosh y \times  \\
&&\frac{\varepsilon_{c,d,r}\left(\frac{\mu-{m_T}_i \cosh y}{T}\right) \; W_0\left[B\left(1-\frac{\mu-{m_T}_i \cosh y}{r T}\right)^{\frac{1}{d}}\right]}{(1-c)\left[1\pm\varepsilon_{c,d,r}\left(\frac{\mu-{m_T}_i \cosh y}{T}\right)\right] \left(r T+ {m_T}_i \cosh y-\mu \right) \left(1+W_0\left[B\left(1-\frac{\mu-{m_T}_i \cosh y}{r T}\right)^{\frac{1}{d}}\right]\right)}. \nonumber
\end{eqnarray}
At mid-rapidity, i.e. $y=0$, and $\mu=0$,
\begin{eqnarray}
\frac{1}{2 \pi}\frac{d^2 N}{dy m_T dm_T}&=&\pm \frac{V}{8 \pi^3} \sum_i g_i T {m_T}_i \times \nonumber \\ &&\frac{\varepsilon_{c,d,r}\left(\frac{-{m_T}_i}{T}\right) \; W_0\left[B\left(1-\frac{(-{m_T}_i)}{r T}\right)^{\frac{1}{d}}\right]}{(1-c)\left[1\pm\varepsilon_{c,d,r}\left(\frac{-{m_T}_i}{T}\right)\right] \left(r T+{m_T}_i\right) \left(1+W_0\left[B\left(1-\frac{(-{m_T}_i)}{r T}\right)^{\frac{1}{d}}\right]\right)}.
\end{eqnarray}
The transverse momentum distribution reads
\begin{eqnarray}
\left.\frac{1}{2 \pi}\frac{d^2N}{p_T dy dp_T}\right|_{y=0}&=&\pm\frac{V}{8 \pi^3} \sum_i g_i T {m_T}_i \times \nonumber \\ &&\frac{\varepsilon_{c,d,r}\left(\frac{-{m_T}_i}{T}\right) \; W_0\left[B\left(1-\frac{(-{m_T}_i)}{r T}\right)^{\frac{1}{d}}\right]}{(1-c)\left[1\pm\varepsilon_{c,d,r}\left(\frac{-{m_T}_i}{T}\right)\right] \left(r T+{m_T}_i\right) \left(1+W_0\left[B\left(1-\frac{(-{m_T}_i)}{r T}\right)^{\frac{1}{d}}\right]\right)}. \hspace*{8mm}
 \label{transmomentum}
\end{eqnarray}
At mid-rapidity, i.e. $y=0$, but $\mu\neq 0$
\begin{eqnarray}
\left.\frac{1}{2 \pi}\frac{d^2N}{p_T dy dp_T}\right|_{y=0}&=&\pm\frac{V}{8 \pi^3} \sum_i g_i T {m_T}_i \times \nonumber \\ &&\frac{\varepsilon_{c,d,r}\left(\frac{\mu-{m_T}_i}{T}\right) \; W_0\left[B\left(1-\frac{(\mu-{m_T}_i)}{r T}\right)^{\frac{1}{d}}\right]}{(1-c)\left[1\pm\varepsilon_{c,d,r}\left(\frac{\mu-{m_T}_i}{T}\right)\right] \left(r T-\mu+{m_T}_i\right) \left(1+W_0\left[B\left(1-\frac{(\mu-{m_T}_i)}{r T}\right)^{\frac{1}{d}}\right]\right)}. \hspace*{8mm}
 \label{transmomentum1}
\end{eqnarray}

We can now fit various transverse momentum distributions measured in different experiments, i.e. different types of collisions and different collision energies, rapidity, centrality, etc. by using Eq. (\ref{transmomentum}) and Eq. (\ref{transmomentum1}) at mid-rapidity.  Expression (\ref{transmomentum1}) takes into consideration vanishing and finite baryon chemical potentials \cite{Tawfik:2014eba,Tawfik:2013bza}.

\begin{figure}
\begin{center}
\includegraphics[width=5.5cm,angle=-90]{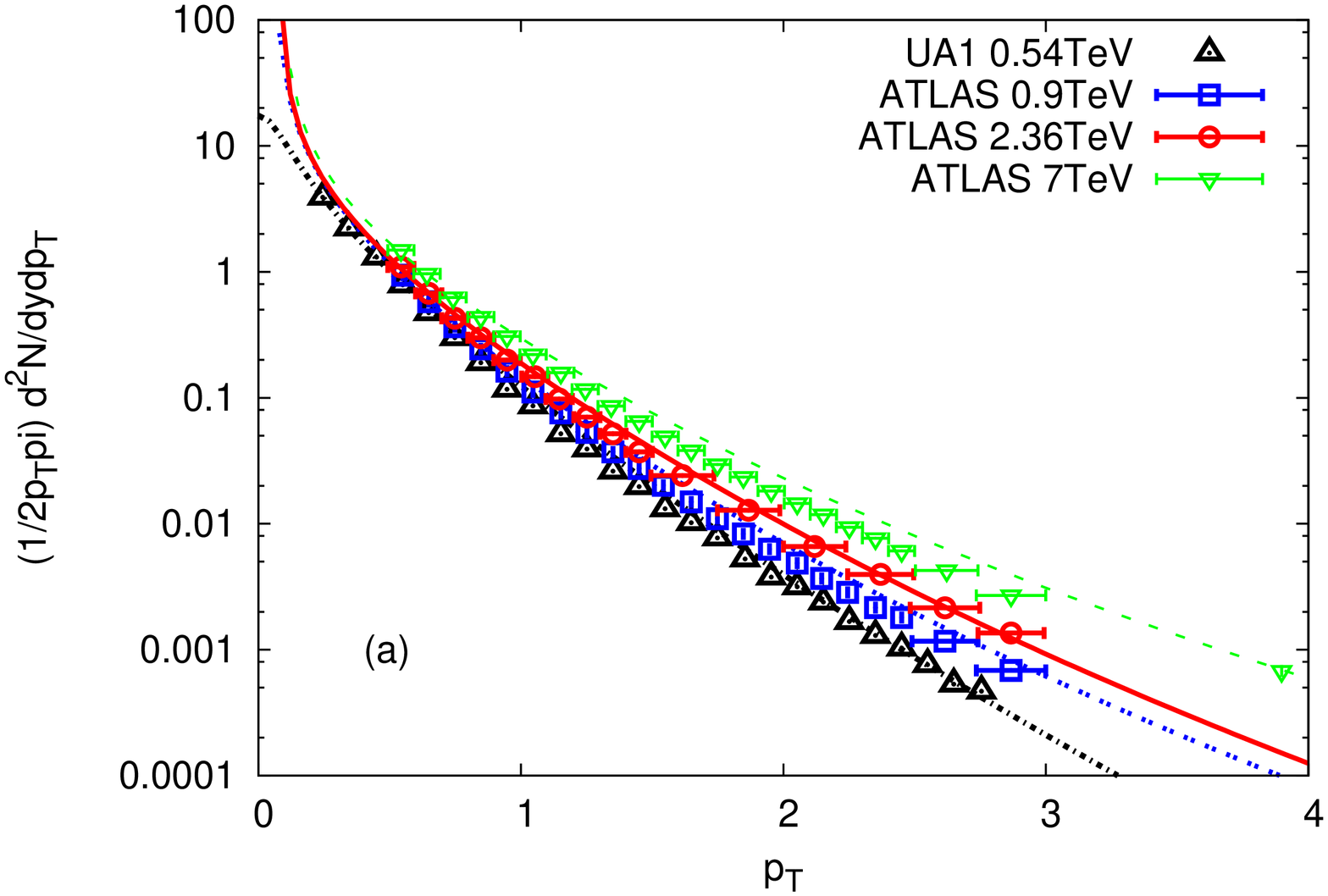} \\
\includegraphics[width=8.cm,angle=-00]{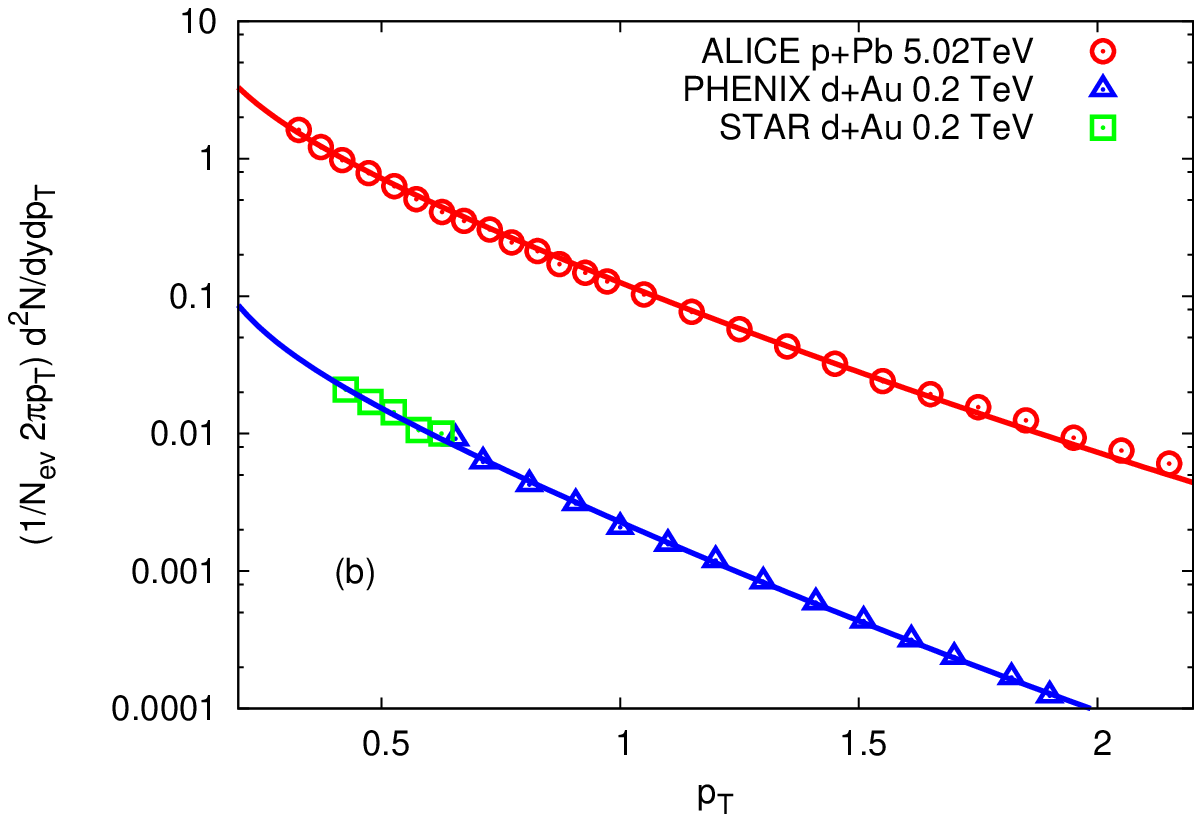}\\
\includegraphics[width=8.cm,angle=-00]{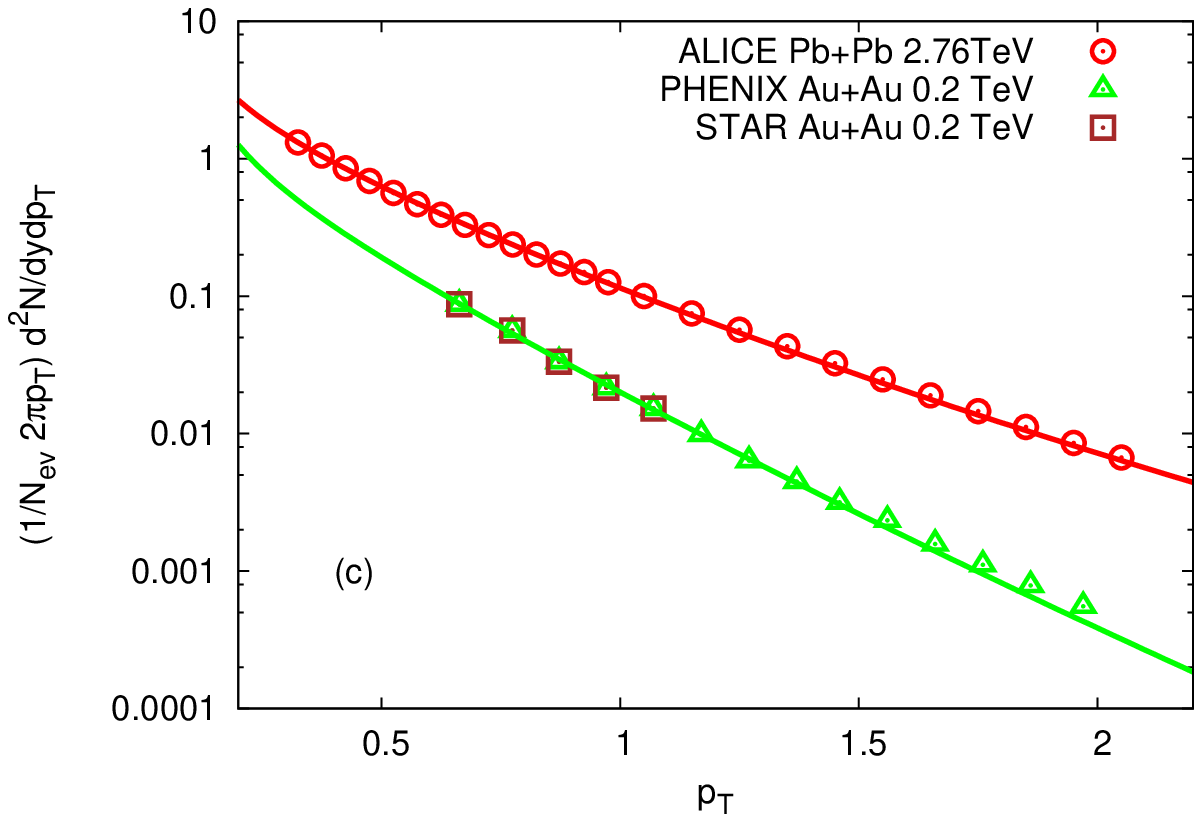}
\figcaption{Transverse momentum distributions for the charged particles $\mathtt{K}^+, \mathtt{\pi}^+$ and $\mathtt{p}$ and their antiparticles measured in the experiments UA1 \cite{UA1}, PHENIX \cite{Adare:2013esx}, STAR \cite{Abelev:2008ab}, ATLAS \cite{reFFF4} and ALICE \cite{Abelev:2013haa,ALICE} (symbols) are compared with calculations from the genetic nonextensive statistical approach (curves). The resulting fit parameters are listed in Tab. \ref{table:1}. The pp, pA and AA results are separately depicted in top, middle and bottom panels, respectively.}
\label{fig:pT}
\end{center}
\end{figure}

\begin{figure}
\begin{center}
\includegraphics[width=6.cm,angle=-90]{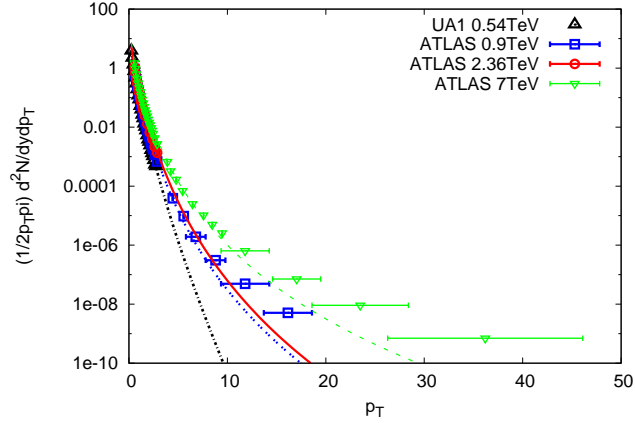} \\
\figcaption{The same as in Fig. \ref{fig:pT} but here including the largest $p_T$-region. }
\label{fig:highpT}
\end{center}
\end{figure}

Figure \ref{fig:pT} depicts the transverse momentum distributions for the charged particles $\mathtt{K}^+, \mathtt{\pi}^+$ and $\mathtt{p}$ and their antiparticles measured in different types of collisions at various collision energies; p+p collisions at $0.54$ \cite{UA1}, $0.9$, $2.36$ and  $7~$TeV \cite{reFFF4}, p+Pb collisions at $5.02~$TeV \cite{Abelev:2013haa}, Pb+Pb collisions at $2.76~$TeV \cite{ALICE}, d+Au collisions and Au+Au collisions at $0.2~$TeV \cite{Adare:2013esx,Abelev:2008ab}. This set of measured particles combines momentum spectra of six charged particles. Here, we zoom out the smallest $p_T$ region. The entire $p_T$ spectra are illustrated in Fig. \ref{fig:highpT}. All these distributions are fitted to the generic nonextensive statistical approach, Eq. (\ref{transmomentum}). The resulting fit parameters are given in Tab \ref{table:1}. For a better comparison, the p+p, p+A and A+A results are separately depicted in top, middle and bottom panels, respectively.

The entire $p_T$-range is presented in Fig. \ref{fig:highpT}. It is noteworthy highlighting that the statistical fit was indeed performed over the complete range of $p_T$. Figure \ref{fig:pT} is there to zoom out a smaller $p_T$-window. In the top panel (a), we can compare between the pp transverse momentum distributions at different collision energies. There are obvious trends that increasing the collision energies increases $d^2 N/dy dp_T$. Furthermore, at a given collision energy, $d^2 N/dy dp_T$ rapidly exponentially decreases with increasing  $p_T$.

All measurements are performed at mid-rapidity and we assume a vanishing chemical potential. The quality of the statistical fit looks very excellent, see Tab. \ref{table:1}. In p+p collisions at $0.546~$TeV, we find that $c\simeq 1$, while $d\simeq 1.35$. Similar values are also obtained in p+A and A+A collisions, panels (b) and (c). These two values of the equivalence class $(c,d)=(1,d)$ are associated with asymptotically stable systems. At  the resulting $d$, which is positive,  asymptotically stable systems - in tern - are associated with stretched exponential distributions \cite{1005.0138},
\begin{eqnarray}
\lim_{c\rightarrow 1} \varepsilon_{d,r} (x) &=& \exp\left\{-d\, r \left[\left(1-x/r\right)^{1/d}-1\right]\right\}.
\end{eqnarray}  
That the entropy becomes stretched exponentials \cite{Rref13} leads to the special case 
\begin{eqnarray}
\varepsilon_{d,r} (x) &=& r^{1-d}\, d^{-d}\, \exp(d\, r)\, \Gamma (1+d,d\, r-\ln x) - r\, x.
\end{eqnarray}
It was proved that the numerical analysis becomes impossible without the general case of  Gamma-function, $\Gamma(a,b)=\int_b^{\infty} dt\, t^{a-1}\, \exp(-t)$. 

Furthermore, we observe that, the ATLAS measurements are associated with positive $c$ (close to unity), while $d$ can be approximated as $d\simeq 1$. Such an equivalence class defines a nonextensive entropy which is linearly dependent on or - in other words - composed of extensive entropies, such as Renyi \cite{Shafee2007}. Its optimized entropy is given by probability distribution  ($p_i$), which includes the Lambert function,
\begin{eqnarray}
S_{\beta} &=& \sum_i p_i^{\beta}\, \ln\left(\frac{1}{p_i}\right),
\end{eqnarray}
where $\beta \equiv c$ and $p_i^{\beta}$ is the probability that the entire phase-space is occupied. Although this type of entropy looks analogous to the Tsallis one, $-\sum_i(1-p_i^q)/(1-q)$, it is apparently different, at least, as the latter reaches singularity, at $q=1$, the so-called Shannon limit. Further differences have been elaborated in Ref. \cite{Shafee2007}.

It is worthwhile highlighting that the resulting freezeout temperatures  ($T_{\mathtt{ch}}$) raise with increasing collision energy ($\sqrt{s_{\mathrm{NN}}}$).  Possible explanations of this observation are postponed for a future work. We focus on the discussion of their phenomenologies. The Tsallis nonextensive approach was also utilizing in conducting the same study  \cite{1302.1970}. Accordingly,  there is a remarkable difference between $T_{\mathtt{ch}}$ obtained from our approach and the one deduced from Tsallis-type approach \cite{1302.1970}. The latter are relatively smaller than the earlier one (compare Tab. \ref{fig:pT} of the present work with Table 1 of Ref.  \cite{1302.1970}).

Panel (b) of Fig. \ref{fig:pT} shows the transverse momentum distributions of the three charged particles $\mathtt{K}^+, \mathtt{\pi}^+$ and $\mathtt{p}$ and their antiparticles measured in d+Au collisions at $0.2~$TeV and in p+Pb collisions at $5.02~$TeV. The equivalence class can be constricted from $c\simeq 1$, and $d$. The latter is found almost energy-independent; $1.291\pm0.006$ at $0.2~$TeV to $1.331\pm0.004$ at $5.02~$TeV. The freezeout parameters are very interesting. The resulting freezeout  temperature slightly increases from $170\pm13.038~$MeV at $0.2~$TeV to $175\pm13.228~$MeV at $5.02~$TeV, while the corresponding baryon chemical potential drops from $29~$MeV at $0.2~$TeV  to $0~$MeV at $5.02~$TeV.

Panel (c) of Fig. \ref{fig:pT} presents a comparison between A+A collisions at different collision energies; Au+Au collisions at $0.2~$TeV and Pb+Pb collisions at $2.76~$TeV for the momentum spectra of the three charged particles $\mathtt{K}^+, \mathtt{\pi}^+$ and $\mathtt{p}$ and their antiparticles. Despite the system size difference, we also find that $c\simeq 1$, while $d$ slightly increases from $1.221\pm0.055$ at $0.2~$TeV to $1.357\pm0.023$ at $2.76~$TeV. The freezeout temperature shows an energy-independent behavior; $155\pm12.450~$MeV at $0.2~$TeV to $165\pm12.845~$MeV at $2.76~$TeV. The baryon chemical potential behaves almost similar to p+A collisions, panel (b).

\begin{center}
\tabcaption{\label{table:1} Various parameters deduced for the statistical fit of the transverse momentum calculations based on the generic nonextensive statistical approach (present work) to various measurements, Fig. \ref{fig:pT}.}
\begin{tabular}{|| c | c | c | c | c | c | c | c ||}
 \hline\hline
 Experiment  & $\sqrt{s_{\mathtt{NN}}}$[TeV] & Centrality & $T_{\mathtt{ch}}$[MeV] & $\mu_{\mathtt{b}}$[MeV] & $d$ & $c$ & $\chi^2$ \\
\hline  \hline
 UA1 (p+p) & $0.546$ & $0\%$ & $110 \pm 10.488$ & $0.0$ & $1.35 \pm 0.135$ & $0.999 \pm 0.100$ & $0.384$ \\
  ATLAS (p+p) & $0.9$ & $0\%$ & $140 \pm 11.832$ & $0.0$ & $1.03 \pm 0.103$ & $0.945 \pm 0.095$ & $0.766$ \\
 ATLAS (p+p) & $2.36$ & $0\%$ & $160 \pm 12.649$ & $0.0$ & $1.06 \pm 0.106$ & $0.945 \pm 0.095$ & $1.651$  \\
 ATLAS (p+p) & $7.0$ & $0\%$ & $150 \pm 12.247$ & $0.0$ & $1.05 \pm 0.105$ & $0.930 \pm 0.093$ & $1.115$ \\  \hline
 PHENIX/STAR (d+Au) & $0.2$  & $0-20\%$ & $170 \pm 13.038$ & $29$ & $1.291 \pm 0.006$ & $0.999\pm0.001$ & $1.884$ \\
 ALICE (p+Pb) & $5.02$  & $0-5\%$ & $175 \pm 13.228$ & $0.0$ & $1.333 \pm 0.004$ & $0.999\pm0.001$ & $0.902$ \\  \hline
  PHENIX/STAR (Au+Au)& $0.2$   & $0-10\%$ & $155 \pm 12.450$ & $25$ & $1.221 \pm 0.055$ & $0.999\pm0.001$ & $1.584$ \\
 ALICE (Pb+Pb) & $2.76$  & $0-5\%$ & $165 \pm 12.845$ & $0.0$ & $1.357 \pm 0.023$ & $0.999\pm0.001$ & $0.164$ \\  \hline \hline
\end{tabular}
\vspace{0mm}
\end{center}
\vspace{0mm}

From the fit parameters listed out in Tab. \ref{table:1}, we can summarize that except for ATLAS measurements, $c\simeq 1$, while $d>1$. These values are obtained for different system sizes and at various collision energies. They refer to nonextensivity but not of Tsallis type. For the sake of completeness, we remind again that only at $c=d=1$, the BG extensivity is guaranteed. It has been pointed out \cite{bialas2015} that the Tsallis algebra, which is mainly implemented through replacing the exponential and logarithmic functions by their counterpart expressions in the Tsallis nonextensive approach, can be scaled as power laws. Such a scaling can also be obtained in the so-called statistical cluster-decay. Thus, the good fitting of the transverse momentum distributions by the Tsallis nonextensivity would be misleading \cite{bialas2015}, as the role of the statistical cluster-decay is apparently ignored. In a forthcoming work, we shall estimate the contribution of the statistical cluster-decay to the possible power-laws in the nonextensive fit with a special emphasize to the Tsallis-type nonextensivity. 

In the section that follows, we confront our approach to different particle ratios and yields, which likely can not be scaled as power laws, as such. The resulting (non)extensivity parameters are conjectured to shed light on the statistical nature of the particle production and whether it is an extensive or a nonextensive process and if the latter whether this follows Tsallis approach.

\subsection{Particle ratios and yields}
\label{sec:fittingPR}

\begin{figure}
\begin{center}
\includegraphics[width=5.5cm]{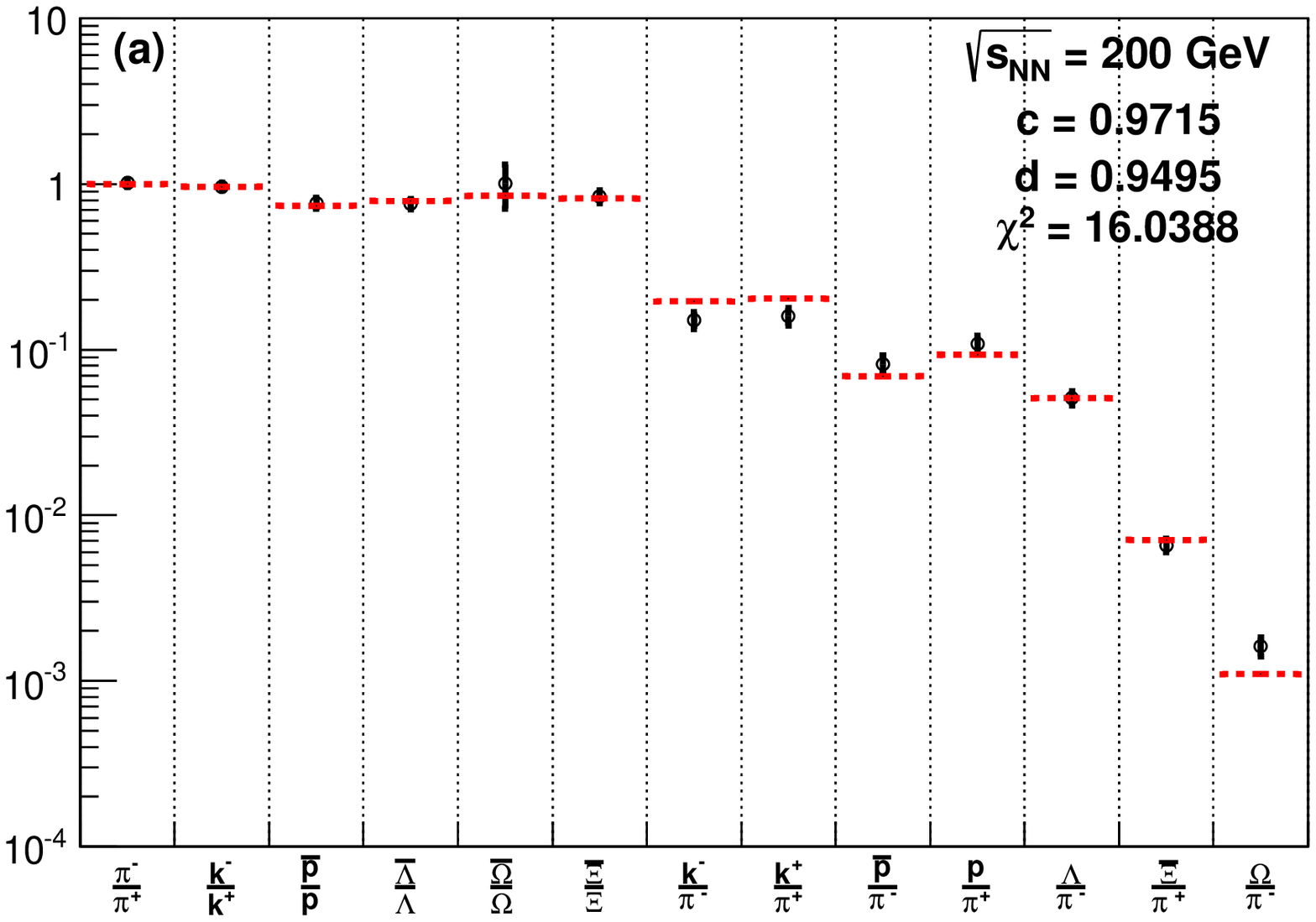}
\includegraphics[width=5.5cm]{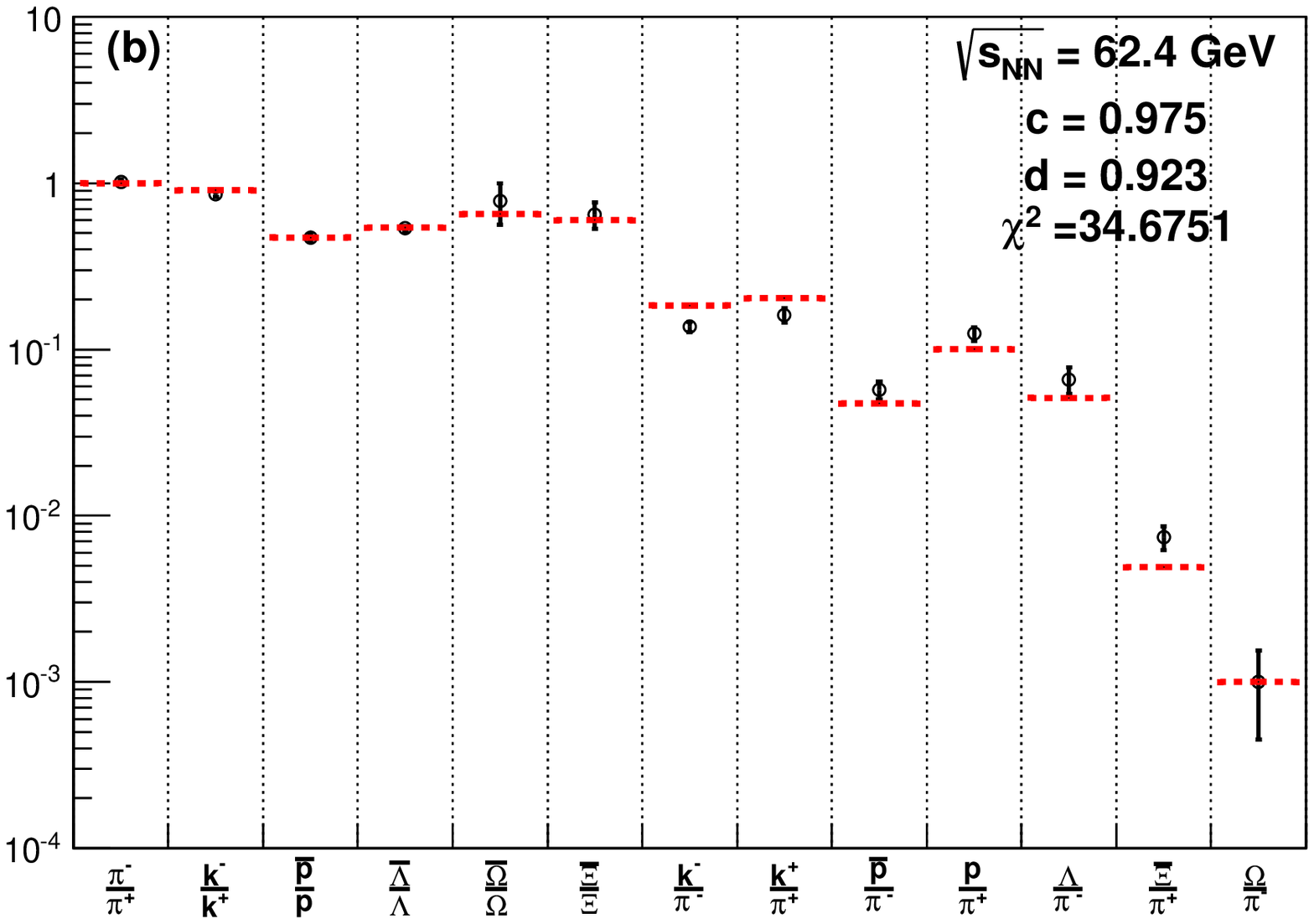}
\includegraphics[width=5.5cm]{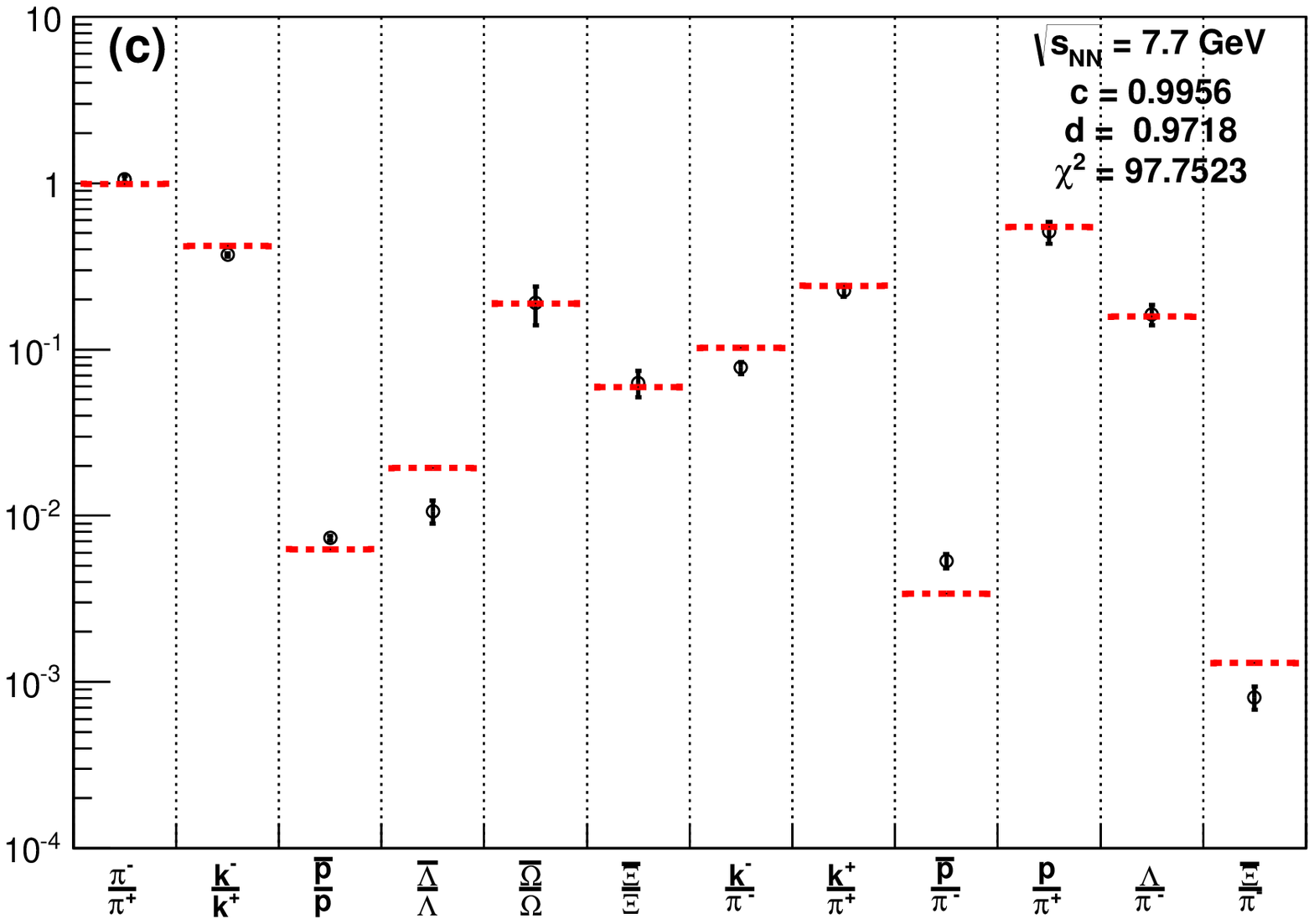}
\figcaption{Panel (a): different particle ratios deduced from the generic axiomatic-nonextensive statistical approach (dashed lines) are compared with the experimental results at $200~$GeV (symbols). Panels (b) and (c) show the same but at $62.4$ and  $7.7~$GeV, respectively. The exponents $c$ and $d$ and $\chi^2$ are given in top right corner. \label{fig:prnonext}}
\end{center}
\end{figure}

The particle ratios $\pi^-/\pi^+$, $K^-/K^+$, $\bar{p}/p$, $\bar{\Lambda}/\Lambda$, $\bar{\Omega}/\Omega$, $\bar{\Xi}/\Xi$, $K^-/\pi^-$, $K^+/\pi^+$, $\bar{p}/\pi^-$, $p/\pi^+$, $\Lambda/\pi^-$, $\Omega/\pi^-$, and $\bar{\Xi}/\pi^+$ measured in Au+Au collisions in the STAR experiment at energies $200~$GeV, $62.4~$GeV, and $7.7~$GeV are statistically fitted by means of the HRG model, in which the generic nonextensive statistical approach is implemented. The number density can be derived from the partition function and accordingly the particle ratios can be determined. 

We take into account all possible decays into the particle of interest and their branching ratios. $\mu$, $T$, $c$, and $d$ are taken as free parameters. The strangeness chemical potential is calculated at each value assigned to $\mu$, $T$, $c$, and $d$ so that an overall strangeness conservation is guaranteed. The results depicted in Fig. \ref{fig:prnonext} represent best agreement between measurements and calculations, i.e. at the smallest $\chi^2$ value. The results at $200~$GeV [panel (a)],  $62.4~$GeV [panel (b)] and $7.7~$GeV [panel (c)] are depicted in Fig. \ref{fig:prnonext}. We observe that the quality of the fits weakens with the collision energies; $\chi^2/\mathtt{dof}=1.105$, $\chi^2/\mathtt{dof}=2.771$ and $\chi^2/\mathtt{dof}=8.146$ at $200~$GeV,  $62.4~$GeV and $7.7~$GeV, respectively. 

As mentioned, our fits for the transverse momentum distributions and their debatable interpretation whether the resulting parameters were due to power laws are stemming from Tsallis-algebra should be a subject of a future work. Here, we have analysed another thermodynamic quantity, the particle ratios, which are conjectured to highlight the (non)extensivity, as they are not to be scaled as power laws. In light of this assumption, we can discuss the resulting parameters and their physical meanings. 

The resulting freezeout temperature ($T_{\mathtt{ch}}$) and baryon chemical potential ($\mu_{\mathtt{b}}$) can be summarized as follows.
\begin{itemize}
\item at $200~$GeV, $T_{\mathtt{ch}}=148.05 \pm 12.168$MeV and $\mu_{\mathtt{b}}=23.94 \pm 4.89$MeV,
\item at $62.4~$GeV, $T_{\mathtt{ch}}=179.13 \pm 13.384$MeV and $\mu_{\mathtt{b}}=57.33 \pm 7.57$MeV, and
\item at $7.7~$GeV, $T_{\mathtt{ch}}=145.32 \pm 12.055$MeV and $\mu_{\mathtt{b}}=384.3 \pm 19.6$MeV,
\end{itemize}
which are obviously very compatible with the ones deduced from BG statistics \cite{Tawfik:2013bza}. The equivalence class reads
\begin{itemize}
\item At $200~$GeV, $c=0.971 \pm 0.097$ and $d=0.949 \pm 0.09$,
\item at  $62.4~$GeV, $c=0.975 \pm 0.098$ and $d=0.923 \pm 0.091$ and
\item at $7.7~$GeV, $c=0.995 \pm 0.1$ and $d=0.972 \pm 0.097$.
\end{itemize}
These can be approximated as both $c$ and $d$ lie below unity referring neither to BG extensivity nor to Tsallis nonextensivity.  This conclusion still needs a further analysis at other collision energies. We shall devote a future work to examine such a relation.

\begin{figure}
\begin{center}
\includegraphics[width=5.5cm]{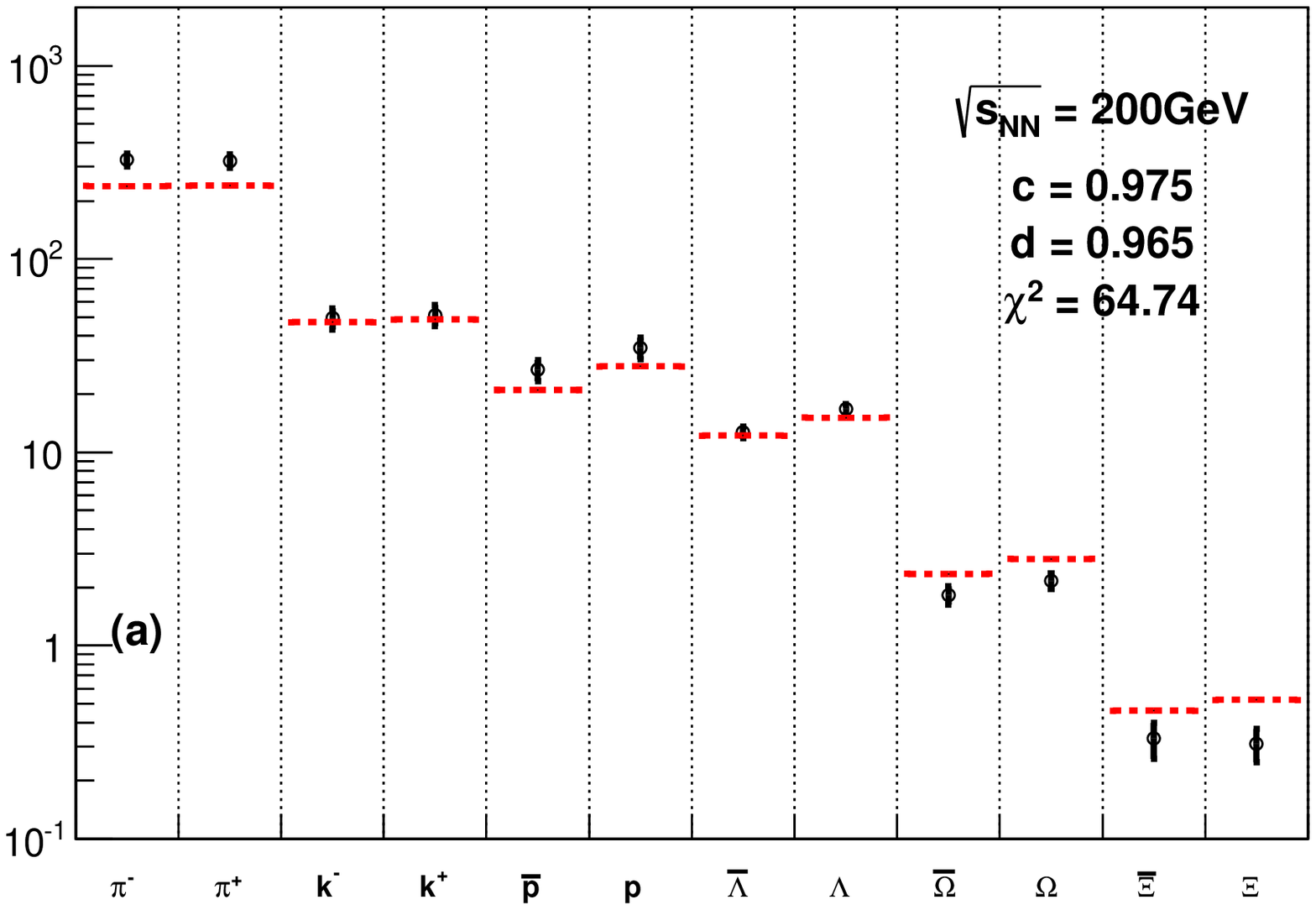}
\includegraphics[width=5.5cm]{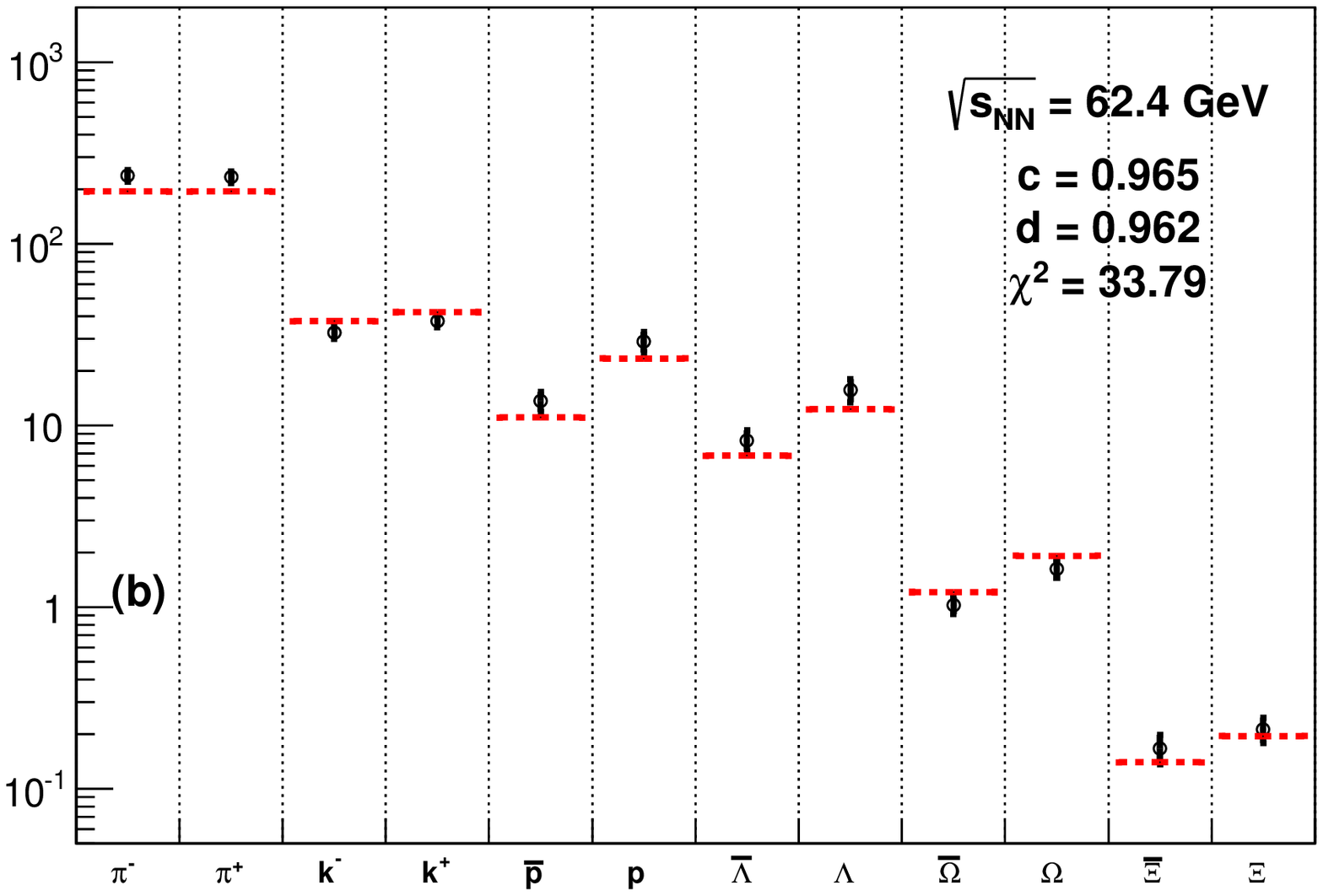}
\includegraphics[width=5.5cm]{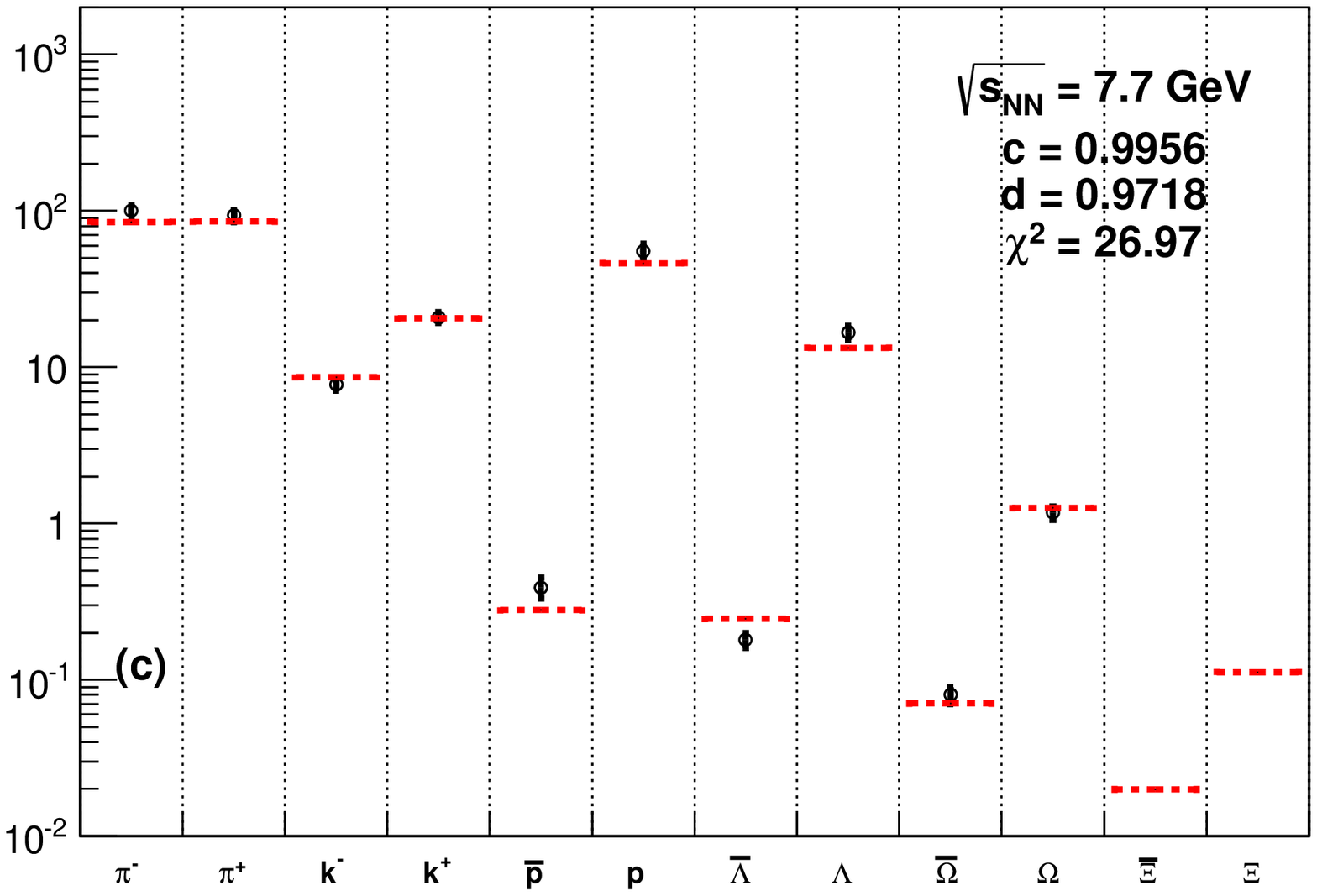}
\figcaption{The same as in Fig. \ref{fig:prnonext} but here for the particle yields. \label{fig:pyieldnonext}}
\end{center}
\end{figure}

\begin{table}
\begin{center}
\tabcaption{\label{table:2} The freezeout parameters and equivalence class deduced for the statistical fit of the generic nonextensive statistical approach (present work) to the experimental results on particle yields, Fig. \ref{fig:pyieldnonext}.}
\footnotesize
\begin{tabular}{||c | c | c | c | c | c | c | c||}
 \hline\hline
 $\sqrt{s_{\mathtt{NN}}}$ [GeV] & Centrality & $T_{\mathtt{ch}}$ [MeV] & $\mu_{\mathtt{b}}$ [MeV]& d & c & R [fm] & $\chi^2$\\
\hline \hline
 $200$ & $0\%$ & $160.54 \pm 12.67$ & $25.20 \pm 5.02$  & $0.965 \pm 0.097$ & $0.975 \pm 0.098$ & $2.044 \pm 1.126$ & $5.395$ \\
 $62.4$ & $0\%$ & $170.00 \pm 13.04$ & $57.83 \pm 7.60$ & $0.962 \pm 0.096$ & $0.965 \pm 0.097$ & $1.796 \pm 1.056$ & $2.816$ \\
 $7.7$ & $0\%$ & $147.31 \pm 12.14$ & $388.08 \pm 19.7$ & $0.972 \pm 0.097$ & $0.995 \pm 0.101$ & $1.848 \pm 1.071$ & $2.701$\\ \hline \hline
\end{tabular}
\end{center}
\end{table}

In Fig. \ref{fig:pyieldnonext}, we present fits for various particle yields, $\mathrm{\pi}^-$, $\mathrm{\pi}^+$, $\mathrm{K}^-$, $\mathrm{K}^+$, $\bar{\mathrm{p}}$, $\mathrm{p}$,  $\bar{\mathrm{\Lambda}}$, $\mathrm{\Lambda}$, $\bar{\mathrm{\Omega}}$, $\mathrm{\Omega}$, $\bar{\mathrm{\Xi}}$, and $\mathrm{\Xi}$ measured in Au+Au collisions in the STAR experiment (symbols) at $200~$GeV, $62.4~$GeV, and $7.7~$GeV, respectively, are fitted to calculations from the HRG model with generic nonextensive statistical approach (dashed lines). The equivalence class is given in the top-right corners of each graph. Together with the resulting freezeout parameters, they are listen out in Tab. \ref{fig:pyieldnonext}.

In summary, we have found that
\begin{itemize}
\item for $p_T$ spectra except ATLAS, $c\simeq 1$ and $d>1$. The nonextensive entropy is associated to stretched exponentials, where Lambert function reaches its asymptotic stability.
\item for ATLAS $p_T$ spectra, $c<1$, while  $d\simeq 1$. The nonextensive entropy is linearly composed of extensive entropies. The reasons why ATLAS measurements look different from others shall be discussed in a future work. 
\item for particle ratios and yields, $c<1$ and $d<1$. This is known as $(c,d)$-entropy, where $d>0$ and Lambert functions $W_0$ exponentially raise.
\end{itemize}

\section{Conclusions and outlook}
\label{sec:cncl}

In this work, we present a systematic study for the thermodynamic self-consistency of the generic axiomatic-nonextensive statistical approach, which is characterized by two asymptotic properties. To each of them, a scaling function is assigned. These scaling functions are estimated by exponents $c$ and $d$ for first and second property, respectively. In the thermodynamic limit of grand-canonical ensembles of classical (Boltzmann) and quantum gas (Fermi-Dirac and Bose-Einstein statistics) of a gas composed on various hadron resonances. We start with first and second laws of thermodynamic which characterize the statistical system of interest. The thermodynamic properties of that system are determined and confirmed that both laws of thermodynamic are fully verified. We have proved that the definitions of temperature, number density, chemical potential and entropy density within the generic axiomatic-nonextensive classical and quantum statistical approach lead to expressions which satisfy the laws of thermodynamics.

The second part of this script introduces implementations of the generic axiomatic-nonextensive  statistical approach to high-energy physics. We start with the transverse momentum distributions measured in central collisions, in different system sizes and at various collision energies. We found that, $c\simeq 1$ and $d>1$, which describe nonextensive entropy associated to stretched exponentials, in which the Lambert function behaves, asymptotically. For the ATLAS $p_T$ spectra, $c<1$, while  $d\simeq 1$. This is nonextensive entropy, which is linearly involves extensive entropies, such as Renyi. All these values differ from the equivalence class characterizing BG- and Tsallis-statistics. The role of the statistical cluster-decay was also highlighted. Their contributions to the possible power-laws should be first evaluated, so that the ones stemming from the Tsallis-type nonextensivity can be determined.

We also calculated different particle ratios and yields, which likely are not to be scaled as power laws. From the resulting exponents $c$ and $d$, we can judge whether the particle production is a (non)extensive process. We found that $c<1$ and $d<1$ referring to $(c,d)$-entropy, where the Lambert functions raise,  exponentially. It is apparent that this finding points out to neither BG extensivity nor to Tsallis nonextensivity.

From the statistical fit of both sets of experimental results ($p_T$ and particle ratios and yields), the resulting freezeout parameters, $T_{\mathtt{ch}}$ and $\mu_{\mathtt{b}}$, are fairly compatible with the ones deduced from BG statistics. We believe that the statistical properties either extensivity or nonextensivity, shouldn't be related to intensive or extensive thermodynamic quantities, such as temperature and baryon chemical potential. These are strongly associated with the proposed equivalence class ($c,d$). The latter characterizes BG, or Tsallis or even genetic statistics, where  ($1,1$), or ($c,0$) or ($c,d$), respectively.

Finally, we conclude that the proposed generic nonextensive statistical approach is thermodynamically self-consistent and greatly able to reveal the statistical nature of various processes taking place in high-energy collisions, such as transverse momentum spectra and particle ratios and yields. 

Many authors endorse the assumption that the Tsallis-type nonextensivity can be originated to fluctuations, correlations and inter-particle interactions. We shortly discussed discussed on all these in the introduction. Wilk {\it et al.} \cite{Wilk2000} proposed that the nonextensive parameter ($q$) would be related to the fluctuations in the inverse temperature. It is believed that the temperature reflects impacts of the nonextensivity. Since this interesting topic lie out of the scope of the present paper, we plan in a future work to examine the possibilities that the equivalence class ($c,d$) instead of the temperature might signature any of the physical processes, fluctuations, correlations and inter-particle interactions, if not all.


\clearpage
\end{CJK*}
\end{document}